\documentclass{iopart}
\usepackage{iopams} 
\usepackage{setstack}
\usepackage{geometry}
\usepackage{color}
\usepackage{amssymb}
\usepackage{mathrsfs}
\usepackage{verbatim}
\pdfoutput=1

\usepackage{graphicx,epsfig}
\usepackage{bm}

\usepackage{braket}

\newcommand{\be}{\begin{equation}}
\newcommand{\ee}{\end{equation}}
\newcommand{\bea}{\begin{eqnarray}}
\newcommand{\eea}{\end{eqnarray}}

\begin{document}
\title[Semi-classical theory for quantum quenches in the O(3) non-linear sigma model]{Semi-classical theory for quantum quenches in the O(3) non-linear sigma model} 
\author{Stefano Evangelisti$^{1,2}$} 
\address{$^1$ The Rudolf Peierls Centre for Theoretical Physics, University of Oxford -  Oxford OX1 3NP, United Kingdom}
\address{$^2$ Department of Physics, University of Bologna and INFN, Via Irnerio 46, 40136, Bologna, Italy}
\begin{abstract}
We use the semi-classical approach to study the non-equilibrium dynamics of the O(3) non-linear sigma model. For a class of quenches defined in the text, we obtain the order-parameter dynamical correlator in the thermodynamic limit. In particular we predict quench-dependent  relaxation times and correlation lengths. The approach developed for the O(3) non-linear sigma model can also be applied to the transverse field Ising chain, where the semi-classical results can be directly compared to both the exact and the numerical ones, revealing the limits of the method. 
\end{abstract}

\maketitle

\section{Introduction}%
The out-of-equilibrium physics of many-body systems has attracted a lot of interest in recent years, not least because of the experimental realizations of
ultracold atomic gases in optical lattices \cite{Greiner,Kinoshita,Hofferberth,Trotzky,Cheneau,Gring}.  
These experiments have observed the dynamics of many-body systems on a long time scale after a quantum quench, finding essentially unitary time-evolution.  In three-dimensional systems fast relaxation towards a thermal steady state has been observed. On the contrary, in quasi-one-dimensional systems the relaxation process is normally much slower and leads to a peculiar non-thermal stationary state \cite{Kinoshita}. These results have led to a huge theoretical push \cite{Hofferberth,Trotzky,Cheneau,Gring,Polkovnikov,Rigol1,Rigol2,Calabrese1,Cazalilla1,Cazalilla2,Cazalilla3,Cazalilla4,Barthel,Rossini1,Rossini2,Fioretto1,Biroli,Banulus,Gogolin,Rigol3,Mossel,Caux,Manmana} to address fundamental questions such as whether there is an asymptotic stationary state, and, if it exists, which {\em ensemble} characterizes it. The belief is that observables of non-integrable systems effectively thermalize, which implies that their stationary state is characterized by a thermal Gibbs ensemble. Numerical works on non-integrable systems confirm this expectation, even if some contradictory results point out that some issues have not been completely understood yet \cite{Biroli,Banulus,Gogolin,Kollath}.  On the other hand, in integrable systems, because of the existence of {\em local} integrals of motion, the stationary state is expected to be described by a generalized Gibbs ensemble (GGE), where each mode associated with a conseved quantity is characterized by its own temperature. So far results on integrable systems have been focussed on free fermion models, such as the transverse field Ising chain and the quantum XY chain \cite{McCoy1,McCoy2,McCoy3,Igloi4,Sengupta,Fagotti1,Silva1,Silva2,Venuti1,Venuti2,Igloi1,Foini,Igloi2,Calabrese2,Fabian,Calabrese3,Calabrese4,Igloi3}. \\
For the transverse field Ising chain in thermal equilibrium Sachdev and Young first introduced a semi-classical description of the physical properties of the model in terms of ballistically moving quasi-particles \cite{SachdevYoung}. This approach turned out to be incredibly accurate in predicting the temperature dependence of correlation length, relaxation time and in general to compute the order-parameter two point function in the ferromagnetic and paramagnetic phases. For global quenches this technique has been used to describe the dynamics of the transverse field Ising chain and the quantum XY model, with great accuracy \cite{Igloi4,Igloi1,Igloi2,Igloi3}. In these works quantitative features of the relaxation process  have been explained with a quasi-particle picture, which had been introduced before in particular to study the evolution of the entanglement entropy \cite{Fagotti1,Calabrese6,Calabrese5,Eisler}. In practice the quench injects an extensive amount of energy into the system, which creates quasi-particles homogeneously in space, that then move ballistically with constant velocity. Because of momentum conservation, these quasi-particles are created in pairs with opposite momenta and are quantum entangled (within each pair). Their dynamics can be treated classically as long as they do not collide. But since collisions are unavoidable in one dimension,  every scattering process has to be treated quantum mechanically. We shall use this semi-classical approach to calculate the dynamical correlation functions analytically for the O(3) non-linear sigma model. The dynamics of this model have already been studied at finite temperature in equilibrium \cite{Rapp2,Sachdev}, where the order-parameter correlation function shows a {\em universal} form. Here we shall study its behaviour after a quantum quench, preparing the system in a state which is not an eigenstate of the Hamiltonian. In addition a section will be dedicated to the case of the transverse field Ising chain, whose order-parameter two-point correlation function can be derived straightforwardly from the general one obtained for the O(3) non-linear sigma model.\\
The paper is organized as follows: first we introduce the model and the general form of the initial states we shall consider throughout the whole article, 
before introducing and commenting the main result of the paper, concerning the O(3) non-linear sigma model. Then we describe the basic ideas behind the semi-classical technique and see in detail how to apply them in the present case.
\section{The Model}
\label{s:model}%
Let us start by considering the one-dimensional $O(3)$ quantum rotor chain, which is described by the following Hamiltonian:
\begin{equation}
\label{eq:1}
\hat{H}^{\rm rotor}=\frac{Jg}{2}\sum_{i}\hat{L}_{i}^{2}-J\sum_{i}\hat{n}_{i}\hat{n}_{i+1},
\end{equation}
where $\hat{n}_{i}$ is the position operator of the rotor on site $i=1,\dots,L$ with the constraint $\hat{n}_{i}^{2}=1$\,$\forall_{i}$, and $\hat{L}_{i}=\hat{n}_{i}\times \hat{p}_{i}$ is the angular momentum operator. $J$ is an overall energy scale and $g$ is a positive coupling constant. The operators which appear in the Hamiltonian satisfy the usual commutation relations $[\hat{L}_{i}^{\alpha},\hat{L}_{i}^{\beta}]=i\epsilon_{\alpha\beta\gamma}\hat{L}_{i}^{\gamma}$ and $[\hat{L}_{i}^{\alpha},\hat{n}_{i}^{\beta}]=i\epsilon_{\alpha\beta\gamma}\hat{n}_{i}^{\gamma}$, where $\alpha,\beta,\gamma$ represent the three spatial directions. 
The continuum limit of this lattice model is the  O(3) non-linear sigma model (${\rm nl\sigma m}$), whose Lagrangian density reads:
\begin{equation}
\label{eq:1quatris}
\mathcal{L}={\displaystyle\frac{1}{2\tilde{g}}(\partial_{\mu}\tilde{n}_{i})^{2}},\quad \tilde{n}_{i}^{2}=1,
\end{equation}
where $\tilde{n}_{i}=\tilde{n}_{i}(r,t)$ are three scalar fields and $\tilde{g}$ is a (bare) coupling constant. Here we have already set the maximal propagation velocity $c=1$.  This model is O(3)-symmetric, renormalizable and asymptotically free, and it has three massive particles in the O(3)-multiplet.  The exact $\mathcal{S}$-matrix of this model is known \cite{Zamolodchikov}, with any scattering event involving no particle production and the general $n$-particle $\mathcal{S}$-matrix factorizes into a product of two-particle amplitudes. 
It is worth noticing that the O(3) non-linear sigma model also provides a description of the low-energy excitations of the one-dimensional antiferromagnetic $S=1$ Heisenberg chain \cite{Haldane}, whose dynamical correlation functions' lineshape can be measured experimentally. 
In the $g\gg 1$ limit, in the ground state of Hamiltonian (\ref{eq:1}) all rotors must be in the $L_{i}^{2}=0$ state to minimize the kinetic energy.  The low-energy excitations above the ground state form a triplet with quantum numbers $L_{i}^{z}=(-1,0,1)$.  It is worth remarking that the structure of the low-energy spectrum is the same for any $g > 0$ and the system has a gap $\Delta(g)$. A finite gap is a necessary condition to apply a semiclassical approximation, as we shall see in the next section (for a deeper introduction to the semi-classical method and its range of applicability see \cite{Sachdev} and references therein).\\
In this article we study the dynamical correlator of the order parameter $\tilde{n}^{z}$ after having prepared the system in a squeezed coherent state, namely\footnote{Here for a matter of simplicity we ignore the existence of zero-rapidity terms in the definition of $|\psi\rangle$. In addition it is known that an integrable boundary state as written in equation (\ref{eq:2}) is typically not normalizable, because the amplitude $K$ does not go to zero for large momenta. 
Therefore one has to introduce an extrapolation time in order to make the norm of the state finite, as was done for instance in \cite{Fioretto1}.}:
\begin{equation}
\label{eq:2}
|\psi\rangle = \exp{\left ({\displaystyle\sum_{a, b=0,\pm 1}}\int_0^{\infty}\frac{dk}{2\pi}\, K^{ab}(k)Z^{\dag}_{a}(-k)Z^{\dag}_{b}(k) \right)}|0\rangle,
\end{equation}
where $|0\rangle$ is the ground state of the model, $K^{ab}(k)$ is the amplitude relative to the creation of a pair of particles with equal and opposite momenta, $Z^{\dag}_{a}(k)$ are creation operators of an excitation with quantum number $a=(-1,0,1)$ (the $z$-component of the angular momentum) and momentum $k$.
The main reason for choosing such an initial state comes from its relation with boundary integrable states: as shown by \cite{Calabrese1,Calabrese5} some dynamical problems can be mapped into an equilibrium boundary problem defined in a strip geometry, where the initial state $|\psi\rangle$ acts as a boundary condition. In integrable field theories the most natural boundary states preserve the integrability of the bulk model, or in other words, do not spoil the integrals of motion. These boundary states were originally studied by Ghoshal and Zamolodchikov \cite{Ghoshal1}, and are supposed to capture the universal behaviour of all quantum quenches in integrable models. Furthermore in \cite{Fioretto1} it was shown that any quantum quench of an integrable field theory with this kind of initial states leads to a steady state which is described by a GGE ensemble. Precisely they show that long time limit of the one point function of a local operator can be described by a generalized Gibbs ensemble\footnote{A rigorous proof that the LeClair-Mussardo series work for the one-point functions in the case of a quench has been obtained in \cite{Pozsgay}.}. In short, their result is strong evidence that Rigol {\em et} al.'s conjecture does hold for integrable field theories. \\
The amplitude $K^{ab}(k)$ in expression (\ref{eq:2}) is a regular function which must satisfy a set of constraints that depend on the $\mathcal{S}$-matrix, such as crossing equations, boundary unitarity and boundary Yang Baxter equation. Different solutions of these equations form the set of integrable boundary conditions of the theory.\footnote{For instance for the case of the O(3)  non-linear sigma model with free and fixed boundary conditions these states have been studied in \cite{Ghoshal2}.} \\
The integrable boundary states belong to the class (\ref{eq:2}), but the semi-classical approach applies to the larger class of quenches whose initial states are expressed by a {\em coherent superposition of particle pairs}. In this case the only requirements on the amplitudes $K^{ab}(k)$ are that they must be peaked around $k=0$ and fast enough decreasing functions as $k \to \infty$, so that we can properly define the probabilities in (\ref{eq:res2bis}).\\ 
The order parameter $\tilde{n}^{z}(r)$ in (\ref{eq:1quatris}) may be written as \cite{Sachdev}:
\begin{equation}
\label{eq:2tris}
\tilde{n}^{z}(r,t)\propto (Z^{\dag}_{0}(r,t)+Z_{0}(r,t))+\dots
\end{equation}
where $Z_{0}(r,t)$ is the (time-dependent) Fourier transform of $Z_{0}(k)$ and the ellipses represent multiparticle creation or annihilation terms, which will be considered negligible because they are subdominat in the long-time limit. 
Relation (\ref{eq:2tris}) comes from the observation that the operator $\tilde{n}^{z}(r)$ either creates a quasiparticle ($L_{i}=1$) with $L_{i}^{z}=0$ at position $r$ with some velocity $v$ or destroys one already present in the system, as will be discussed in the section (\ref{s:semi-classics}).\\
The quench protocol is the following: At time $t=0$ we prepare the system in the initial state (\ref{eq:2}) and  for $t>0 $ this state evolves according to the Hamiltonian (\ref{eq:1}):
\begin{equation}
\label{eq:3}
|\psi(t)\rangle= \exp{(-i\hat{H}t)}|\psi\rangle
\end{equation}
On the other side the time-evolution of an operator is written as:
\begin{equation}
\label{eq:4}
\hat{O}(r, t)=\exp{(i\hat{H}t)}\hat{O}(r)\exp{(-i\hat{H}t)}.
\end{equation}
We are going to analyze the two-point correlator:
\begin{equation}
\label{eq:5}
C^{\rm nl\sigma m}(r_{1}, t_{1}; r_{2}, t_{2})=\langle \psi |\tilde{n}^{z}(r_{2}, t_{2})\tilde{n}^{z}(r_{1}, t_{1})| \psi\rangle.
\end{equation}
The autocorrelation function is obtained for $r_1=r_2=r'$, whereas the equal-time correlation function is defined by imposing $t_1=t_2=t'$, even though in this paper the emphasis will be given to the non-equal time case.                  
\section{Summary and discussion of the results}
\label{s:results}%
In this section we summarize the main result of the paper. The semi-classical method is based upon the existence of a small parameter, namely the average density of  excitation pairs with quantum numbers $(a,b)$ $n^{ab}(k)$, in the initial state $|\psi\rangle$:
\begin{equation}
\label{eq:res1}
\frac{\langle\psi|n^{ab}(k)|\psi\rangle}{\langle\psi|\psi\rangle}\equiv f^{ab}_k\approx|K^{ab}(k)|^{2},
\end{equation}
which is valid in the limit ${\rm max}_{k}\,|K^{ab}(k)|^{2}\ll 1$, $\forall a,b$. This limit also defines what we call {\em small quench}, and it is also the first one we take in our semi-classical approach. The explicit computation of (\ref{eq:res1}) will be carried out in appendix A.
In the limit of small quenches we obtain the following expressions for the (non-equal time) two-point function of the order-parameter of the O(3) non-linear sigma model (\ref{eq:5}) in the limit $T\equiv t_2+t_1 \to \infty$:
\begin{equation}
\label{eq:res1bis}
C^{\rm nl\sigma m}(r;t)=C_{\rm propag}(r,t)\,R(r;t),
\end{equation}
where $C_{\rm propag}(r,t)$ corresponds to the coherent propagation of a quasipaticle, while $R(r;t)$ is the relaxation function, which describes the scattering with the excited quasiparticles. By definition $t\equiv t_2-t_1$ and $r=r_2-r_1$ (without any loss of generality we assume both $(r,t)$ to be positive). 
Whilst $C_{\rm propag}(r,t)$ is valid for all $(r,t)$, the classical relaxation which yields the factor $R(r,t)$ makes sense only within the light-cone ($ct>r$).  
Strickly speaking the validity of formula (\ref{eq:res1bis}) requires $ct\gg r$, where multi-particle contributions are subdominant and can be neglected.\footnote{Taking into account multi-particle contributions in the semi-classical method would imply the computation of form-factors at any further order of approximation, which is beyond the purpose of the original semi-classical approach and the present analysis.} In practice we shall see for the subcase of the Ising model that formula (\ref{eq:res1bis})
works for almost everywhere, except for $(r,t)$ very small. We expect this to be the case for the O(3) non-linear sigma-model as well. 
The pre-factor in (\ref{eq:res1bis}) reads:
\begin{equation}
\label{eq:res1tris}
C_{\rm propag}(r,t)={\displaystyle\int\frac{dk}{2\pi}D(k)e^{ikr-i\epsilon(k)t}}
\end{equation}
where $D(k)$ is a form factor and $\epsilon(k)=\sqrt{\Delta^2+(ck)^{2}}$. For the general lattice model of equation (\ref{eq:1}) $D(k)$ is not known. For our continuum theory, neglecting the multi-particle terms in the spectral density \cite{Damle}, we have:
\begin{equation}
\label{eq:res1four}
D(k)={\displaystyle\frac{\mathcal{A}c}{2\epsilon(k)}},
\end{equation}
where $\mathcal{A}$ is a non-universal quasiparticle residue. This gives $C_{\rm propag}(r,t)\propto K_{0}(\Delta\sqrt{r^{2}-t^{2}})$, with $K_0$ the modified Bessel function.
The relaxation function $R(r;t)$ is instead given by:								
\begin{equation}
\label{eq:res2}
\begin{array}{l}
\fl\qquad\qquad\qquad R(r; t)={\displaystyle\int_{-\pi}^{\pi}\frac{d\phi}{2\pi}}\exp{\left(-t(1-\cos\phi){\displaystyle\int_{0}^{\infty}\frac{dk}{\pi}f_k\, v_{k}\Theta[v_{k}t-r]}\right)}\\
\\
\fl\qquad\qquad\qquad\times\,\exp{\left(-r(1-\cos\phi){\displaystyle\int_{0}^{\infty}\frac{dk}{\pi}f_k\,\Theta[r-v_{k}t]}\right)}\cos\left(r\sin\phi{\displaystyle\int_{0}^{\infty}\frac{dk}{\pi}f_k}\right)\\
\\
\fl\qquad\qquad\qquad\times\,{\displaystyle\frac{1+2\cos(\phi)(P^{00}-1)Q+\cos(2\phi)(Q^{2}-P^{00})-P^{00}Q^{2}}{1-2P^{00}\cos(2\phi)+(P^{00})^{2}}}.
\end{array}
\end{equation}
$Q$ is defined by  $Q= {\displaystyle\sum_{\lambda=\pm 1,0}}P^{0\lambda}$, $f_{k}= {\displaystyle\sum_{a,b=0,\pm 1}}f_{k}^{ab}$, $v_k$ is the velocity of the particle with momentum $k$ and $P^{ab}=\int_{0}^{\infty}\frac{dk}{2\pi}P^{ab}(k)$, where
\begin{equation}
\label{eq:res2bis}
P^{ab}(k)=\frac{{\displaystyle f_{k}^{ab}}}{{\displaystyle\sum_{a,b=0, \pm 1}\int_{0}^{\infty}\frac{dk}{2\pi}f_{k}^{ab}}}.
\end{equation}
Formula (\ref{eq:res2}) is the main result of this article, and it describes the long-time behaviour of {\em any} initial state of the form (\ref{eq:2}). Nonetheless if we want the state $|\psi\rangle$ to respect  the O(3)-symmetry 
we must impose $P^{00}=P^{1,-1}=P^{-1,1}=1/3$ ($Q=1/3$), and all other probabilities equal to zero.\\
The most general expression for the relaxation function $R(r; t, T)$ for generic values of the times $(t_1, t_2)$ will be given in the appendix B. 
Formula (\ref{eq:res2}) defines the quench-specific time and length scales:
\be
\label{eq:res3}
\tau^{-1}  =  {\displaystyle\int_{0}^{\infty}\frac{dk}{\pi}v_{k}\,f_k}\qquad\qquad\xi^{-1}  =  {\displaystyle\int_{0}^{\infty}\frac{dk}{\pi}f_k}.
\ee
We also note that the correlator of equation (\ref{eq:res2}) is {\em never} thermal unless one prepares the system at $t=0$ in a thermal inital state. This result confirms the belief that one-dimensional integrable systems relax towards a peculiar non-thermal distribution, namely the generalized Gibbs ensemble (GGE). 

\section{The Semi-classical Theory}\label{s:semi-classics}%
\subsection{From quantum correlators to classical probabilities}

\begin{figure}[t]
   \begin{center}
   \includegraphics[width=4cm]{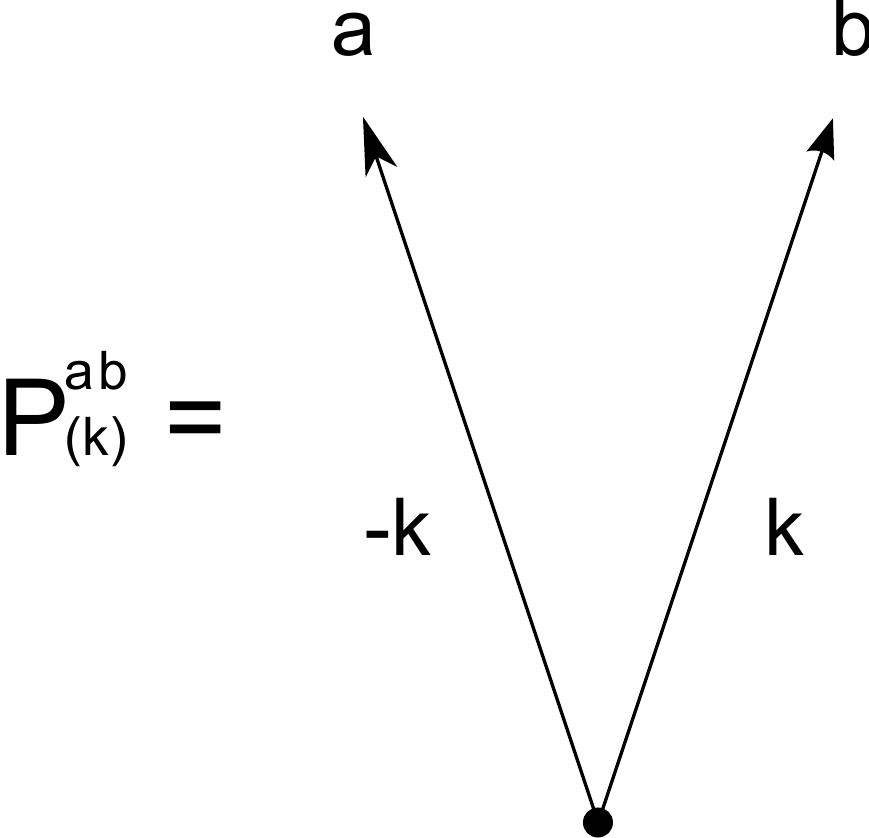}
   \caption{Graphical representation of the probability function $P^{ab}(k)$. }
\label{fig:1}
\end{center}
\end{figure}

\begin{figure}[t]
   \begin{center}
   \includegraphics[width=8cm]{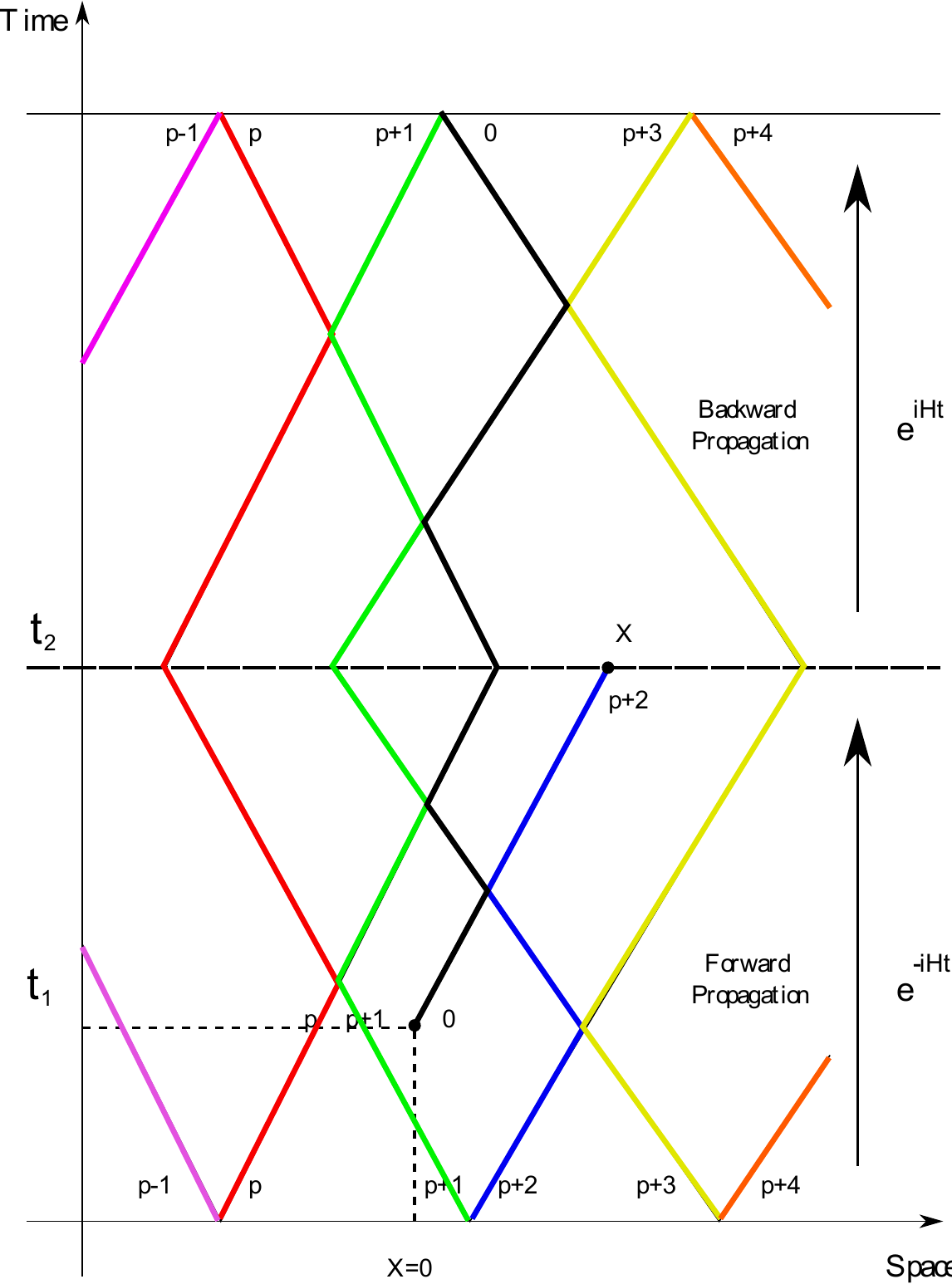}
   \caption{Propagation and scattering of quasiparticles in the semi-classical picture of the $O(3)$ non-linear sigma model. Lines drawn with different colours represent trajectories of different particles. The two states before and after the forward and backward time evolution must be identical, and this imposes constraints on the quantum numbers of the particles. Notice that in this example the particle with label $p+2$ is removed at time $t_2$. }
\label{fig:2}
\end{center}
\end{figure}

The main idea behind the semi-classical approach to the computation of out-of-the-equilibrium correlators is basically encoded in the following representation of the correlator (\ref{eq:5}): 
\begin{equation}
\label{eq:5bis}	
\begin{array}{l}								
C^{\rm nl\sigma m}(r_{1}, t_{1}; r_{2}, t_{2})\approx {\displaystyle\sum_{M}\sum_{\{ \lambda_{\nu}\}}\int\prod_{\nu}dx_{\nu}\prod_{\nu}dk_{\nu}}\\
\times \left[P(\{x_{\nu},k_{\nu},\lambda_{\nu}\})\langle \{ x_{\nu},k_{\nu},\lambda_{\nu}  \} |\tilde{n}^{z}(r_{2}, t_{2})\tilde{n}^{z}(r_{1}, t_{1}) |\{x_{\nu},k_{\nu},\lambda_{\nu}\}\rangle\right],
\end{array}
\end{equation}
where the function $P(\{x_{\nu},k_{\nu},\lambda_{\nu}\})$, is the probability density of having quasiparticles at position $x_{\nu}$ at time $t=0$ with momentum $k_{\nu}$ and quantum number $\lambda_{\nu}$, being $M=N/2$ the total number of quasi-particle pairs in a given initial configuration. The matrix element in the equation represents the value of the correlator once we specify a particular initial configuration for the excitations, which is specified by the set $\{x_{\nu},k_{\nu},\lambda_{\nu}\}$. The summation over $\{ \lambda_{\nu}\}$ averages over all possible quantum numbers of the $N$ particles, which are the three possible values of $L_{i}^{z}=(-1,0,1)$. We will start by assuming the system to be finite with free boundary conditions, and only later will we take the thermodynamic limit.
In spite of this we will assume the system to be translationally invariant, corrispondingly corrections due to the presence of boundaries will be considered negligible (for large enough systems). Therefore the correlator of equation (\ref{eq:5}) becomes a function of $|r_{2}-r_{1}|$,  whilst the same cannot be said for the time dependence, contrary to the equilibrium case.\\
In equation (\ref{eq:5bis}) we have substituted a complicated time-dependent matrix element with a sum over all possibile initial states, which can be represented by pairs of ballistically-moving quasiparticles. This technique has already been used extensively to compute finite-temperature correlators (see for example \cite{Rapp2,Sachdev,Rapp1}), and dynamical correlation functions \cite{Igloi1,Igloi2}. The general idea is the following: each quasiparticle carries a momentum $k_{\nu}$ and a quantum number $\lambda_{\nu}$, and it is created at time $t=0$ together with a partner with equal and opposite momentum. 
If the square modulus of Fourier transform $|\tilde{K}^{ab}(r)|^{2}$ of the amplitude $K^{ab}(k)$ is a fast enough decreasing function as $|r|\to \infty$, the probability of creating these quasiparticles far apart from each other becomes negligible. For this reason in the follow we will always assume quasiparticles to be created in pairs in the same position $x_{\nu}$, where the index $\nu$ labels the quasiparticle pairs. \\
We start by considering this {\em gas} of quasiparticles to be very dilute by tuning the amplitudes $K^{ab}(k)$, therefore the quasiparticle  are entangled only within each single Cooper pair.
This means that the probability of a particular initial state $\{x_{\nu}, k_{\nu}, \lambda_{\nu} \}$ can be factorized in a product of single-pair probabilities:
\begin{equation}
\label{eq:42bis}
\begin{array}{lcl}
P(\{ x_{\nu}, k_{\nu}, \lambda_{\nu} \}) & = &  {\displaystyle \prod_{\nu}}\, P(x_{\nu}, k_{\nu}, \lambda_{\nu}) \\
& = &  {\displaystyle\frac{1}{L^{M}}} {\displaystyle \prod_{\nu} P(k_{\nu}, \lambda_{\nu}) },
\end{array}
\end{equation}
where in the last equality we made use that the system is homogeneous in space (corrections will be present close to the boundaries, but they are negligible in the thermodynamic limit). The quantity $\frac{1}{L} P(k_{\nu}, \lambda_{\nu})$ is the probability for a single pair to be created at a certain position $x_{\nu}$ with momenta $(k_{\nu}, -k_{\nu})$ and quantum numbers $(\lambda'_{\nu}, \lambda''_{\nu})$.
When dealing with Cooper pairs it is convenient to introduce the notation  $P^{\lambda'_{\nu}\lambda''_{\nu}}(k_{\nu})$, whose meaning is:
\begin{equation}
\label{adjointcomment}
P^{\lambda'_{\nu}\lambda''_{\nu}}(k_{\nu})\equiv P(\lambda_{\nu},k_{\nu}),\qquad \lambda_{\nu}\equiv(\lambda'_{\nu},\lambda''_{\nu}),
\end{equation}
where $(\lambda'_{\nu},\lambda''_{\nu})$ are the quantum numbers associated to the two particles in the Cooper pair. This notation is graphically represented in figure (\ref{fig:1}). \\
Without any loss of generality we can also suppose $r_1=0$, thanks to the (quasi)-translational invariance of the system.\\
Calculating the matrix elements for the operator $\tilde{n}^{z}$ (which, as a classical vector would be proportional to the cosine of the azimuthal angle $\theta$) with the first few sherical harmonics, one finds that this operator either creates a quasiparticle with $L_{i}^{z}=\lambda=0$ ($L_{i}=1$) at $r=r_1$ and $t=t_1$ with some velocity $v$ or destroys one already present in the initial state $\{ x_{\nu}, k_{\nu}, \lambda_{\nu} \}$. We assume the latter to be negligible,  because the excitations are very dilute in space and therefore their density is very small (in other words we will use $K^{ab}(k)$ as an expansion parameter). 
This adjoint particle can be created by $\tilde{n}^{z}(r_{1}, t_{1})$ {\em either} inbetween two different quasiparticle pairs {\em or} within two quasiparticle that belong to the same Cooper pair (i.e., two quasiparticles that originated in the same point $x_{\nu}$ at time $t=0$). A possible dynamical scenario is pictured in figure (\ref{fig:2}). 
Due to the collisions with the other excited particles there are only certain configurations where the quantum mechanical overlap in $\langle \psi |\tilde{n}^{z}(r_2,t_2)\tilde{n}^{z}(r_1,t_1)| \psi\rangle$ will be non-zero. Similar to other models the $O(3)$ non-linear sigma model in the long-wavelength limit has  a purely reflective scattering matrix \cite{Damle}:
\begin{equation}
\label{eq:6}
\mathcal{S}_{\lambda_{1} \lambda_{2}}^{\lambda'_{1} \lambda'_{2}} \,\longrightarrow \,(-1)\,\delta_{\lambda_{1}}^{\lambda'_{2}}\delta_{\lambda_{2}}^{\lambda'_{1}}.
\end{equation}
This means that in this limit the particles are impenetrable and the sequence of $\{\lambda_{\nu} \}$ {\em does not} change in time (see figure (\ref{fig:1})). Here we are implicitly assuming that the momentum distribution $f_k$ is a peaked function around zero. 
The operator $\tilde{n}^{z}(r_1,t_1)$ creates a quasiparticle with a probability amplitude $e_{0}(k)$, where the subscript $_{0}$ refers to the quantum number of the created particle and $k$ is its momentum. This particle, together with the other quasiparticles, propagates under the action of $\exp{(-i\hat{H}t)}$ and collides with them (in one-dimensional systems collisions amongst particles can never be ignored). The $\mathcal{S}$-matrix takes on an exchange form (\ref{eq:6}), and therefore particles only exchange their velocities while conserving their internal quantum numbers (see figure (\ref{fig:2})).\\
As a consequence, at any time $t\geq t_1$ precisely one of the particles will have the velocity $v$ of the particle which was created by  $\tilde{n}^{z}(r_1,t_1)$, and will be at position $r_1+v(t-t_1)$. This very particle must be annihilated at time $t_2$ by $\tilde{n}^{z}(r_2,t_2)$, otherwise the final state after the action of $\exp{(i\hat{H}t)}$ will be orthogonal to the initial one.  The probability amplitude that this particle is annihilated is proportional to $(e_{\lambda'}(k))^{*}{\rm e}^{ik(r_2-r_1)}$, where $\lambda'$ is the quantum number of the particle that is removed by $\tilde{n}^{z}(r_2,t_2)$. As we shall see below, $\lambda'$ has to be equal to zero.  This request is automatically guaranteed by another requirement, namely that the internal quantum numbers in the {\em final} state be exactly the same as those of the {\em initial} one. If this does not happen the final state is orthogonal to the initial one, and that particular configuration of quasiparticles does not contribute to the average (\ref{eq:5bis}).\\
Let us now consider the case in which at time $t=0$ the point $r_1=0$ is located inbetween particle with quantum numbers $\lambda_p$ and $\lambda_{p+1}$ (see figure (\ref{fig:2})). Consider now a line connecting the points $(0, 0)$ and $(0, t_1)$ (this line does not correspond to any real particle trajectory).  Define $n' = N'_{+}-N'_{-}$ as the number of intersections between this line and the other lines which correspond to particle trajectories, where $N'_{+}$ are intersections from the right and $N'_{-}$ those from the left. By definition $n'$ is given by:
\begin{equation}
\label{eq:43bis}
n' =  {\displaystyle\sum_{\nu}^{M}}\left\{\Theta[0-x_{\nu}^{1}(t_1)] +\Theta[0-x_{\nu}^{2}(t_1)]
 -    2\Theta[0-x_{\nu}] \right\},
\end{equation}
where the labels $^{(1,2)}$ correspond to the trajectories of two quasiparticle originated at the same point and $\Theta$'s are Heaviside step functions. Then at time $t=t_1$ the operator $\tilde{n}_{z}(0, t_1)$ creates a quasiparticle ($L_{\nu}=1$) with $L_{\nu}^{z}=\lambda_{\nu}=0$ at $r_1=0$ with some momentum $k$. The adjoint particle will be created inbetween particles $\lambda_{p+n'}$ and $\lambda_{p+n'+1}$. Suppose that $n'\geq 0$. The generalization of the following results to the case of negative $n'$ is straightforward. The action of the operator $\tilde{n}_{z}(0, t_1)$ on the initial sequence of quantum numbers may be written as:
\begin{equation}
\label{eq:44bis}
\begin{array}{lcl}
\tilde{n}^{z}(0, t_1) & : &  \{\dots,\lambda_{p+n'}, \lambda_{p+n'+1},\dots \}\\
& \mapsto &  \{\dots,\lambda_{p+n'}, \lambda_{0},  \lambda_{p+n'+1},\dots \}.
\end{array}
\end{equation} 
At time $t=t_2$ the operator $\tilde{n}_{z}(r\equiv r_2, t_2)$ will remove the particle with label $p+n'+n$ from the set of excitations, where $n=N_{+}-N_{-}$ is the difference between the number of intersections from the right and those from the left of the line connecting points $(0, t_1)$ and $(r, t_2)$. This quantity is given by:
\be
\label{eq:45bis}
 n =  {\displaystyle\sum_{\nu}^{M}}\left\{\Theta[r-x_{\nu}^{1}(t_2)] +\Theta[r-x_{\nu}^{2}(t_2)]
     - \Theta[0-x_{\nu}^{1}(t_1)]- \Theta[0-x_{\nu}^{2}(t_1)]\right\}.
\ee
This number tells us which particle is moving along this line at time $t=t_2$, particle that will be removed by the action of   $\tilde{n}_{z}(r, t_2)$:
\begin{equation}
\label{eq:46bis}
\begin{array}{lcl}
\fl\quad\tilde{n}^{z}(r, t_2) & : &  \{\dots,\lambda_{p+n'}, \lambda_0, \lambda_{p+n'+1},\dots, \lambda_{p+n'+n-1}, \lambda_{p+n'+n}, \lambda_{p+n'+n+1},\dots \} \\
& \mapsto &  \{\dots,\lambda_{p+n'}, \lambda_0, \lambda_{p+n'+1},\dots, \lambda_{p+n'+n-1}, \lambda_{p+n'+n+1},\dots \} .
\end{array}
\end{equation} 
Comparing the very last sequence of quantum numbers with the initial one (that at time $t=0$), we end up with the following constraint on the set of quantum numbers:
\begin{equation}
\label{eq:46tris}
\boxed{
\lambda_{0}\equiv\lambda_{p+n'+1}\equiv\lambda_{p+n'+2}\dots\equiv\lambda_{p+n'+n}}
\end{equation}
In practice we have a sequence of $n$ quantum numbers that must all be equal to $\lambda_{0}=0$,  which starts at position $p+n'$. 
To identify the contribution of a particular configuration of quasiparticles we must consider the phase factors.  
 Quasiparticles in the initial state generate a phase factor $\exp{(-it_{2}\sum_{\nu}\epsilon(v_{\nu}))}$ under the action of $\exp{(-it_{2}\hat{H})}$ ($\epsilon(v)$ being the particle energy). This phase factor, however, completely cancels under the action of  $\exp{(it_{2}\hat{H})}$, except for the quasiparticle added by the operator $\tilde{n}^{z}(0, t_1)$. This gives a factor $\exp{(-i(t_2 -t_1)\epsilon(v))}$. Moreover every collision results in a sign change of the many-body wave function, but all these signs cancel under the forward and backward propagations, except those that are associated with collisions with the {\em extra} particle. These give an extra sign $(-1)^{N_{+}+N_{-}}$, which can be conveniently re-expressed as:
\begin{equation}
\label{eq:47bis}
(-1)^{N_{+}+N_{-}}=(-1)^{N_{+}-N_{-}}=(-1)^{n}.
\end{equation}
Collecting all these pieces together we obtain the following expression for the general correlation function (\ref{eq:5}):
\begin{equation}
\label{eq:48bis}
\fl\quad C^{\rm nl\sigma m}(r_1, t_1; r_2, t_2)  = \left( {\displaystyle\int(e_{0}(k)^{*}e_{0}(k)){\rm e}^{-i(k(r_2-r_1)-\epsilon(k)(t_2-t_1))}dk} \right)  \times R(r_1, t_1; r_2, t_2),
\end{equation}
where $R(r_1, t_1; r_2, t_2)$ is the relaxation function, which is given by:
\bea
\label{eq:49bis}
 && R(r_1, t_1; r_2, t_2) = \langle{\displaystyle\sum_{n=-\infty}^{+\infty}\sum_{n'=-\infty}^{+\infty}}(-1)^{n}\delta_{n',\sum_{\nu_1}[\dots]}\delta_{n,\sum_{\nu_2}[\dots]}\nonumber\\
& \, & \times\left[\delta_{\lambda_{0},\lambda_{p+n'+1}}  \delta_{\lambda_{p+n'+1},\lambda_{p+n'+2}}\dots\delta_{\lambda_{p+n'+n-1},\lambda_{p+n'+n}} \right]\rangle_{\{ x_{\nu}, k_{\nu}, \lambda_{\nu}\}},
\eea
where the sums with indices $\nu_1$ and $\nu_2$ are given by equations (\ref{eq:43bis}) and (\ref{eq:45bis}) respectively, while the index $p$ is configuration-dependent, signaling where the adjoint particle is created in each set $\{ x_{\nu}, k_{\nu}, \lambda_{\nu}\}$\footnote{Note that once the set $\{ x_{\nu}, k_{\nu}, \lambda_{\nu}\}$ and the position of the first $\tilde{n}_{z}$ have been specified also the index $p$ is uniquely determined.}. The $(-1)^{n}$ appearing in the previous equation is exactly that introduced in (\ref{eq:47bis}). The average $\langle\dots\rangle_{\{ x_{\nu}, k_{\nu}, \lambda_{\nu}\}}$ reads:
\begin{equation}
\label{eq:50bis}
\langle\dots\rangle_{\{ x_{\nu}, k_{\nu}, \lambda_{\nu}\}}\equiv{\displaystyle\sum_{M}\int_{-L/2}^{L/2}\frac{dx_1}{L}\frac{dx_2}{L}\dots\frac{dx_M}{L}
\sum_{\stackrel{\lambda'_{1}, \lambda''_{1}}{k_1>0}}\sum_{\stackrel{\lambda'_{2}, \lambda''_{2}}{k_2>0}}\dots\sum_{\stackrel{\lambda'_{M}, \lambda''_{M}} {k_M>0}}
\prod_{\nu}^{M}P^{\lambda'_{\nu}\lambda''_{\nu}}(k_{\nu})\dots}.
\end{equation}
From equation (\ref{eq:49bis}) we can immediately see that only those configurations which have a  sequence of $n$ quantum numbers all equal to $\lambda_0\equiv 0$ contribute to the correlator (\ref{eq:5}). 
We first tackle the problem of explicitly computing the average over the quantum numbers $\lambda_\nu$ in equation (\ref{eq:49bis}), keeping the total number of quasi-particle pairs $M$ fixed. In the next section we shall see how to release this constraint. 
Let us consider separately the \emph{four} cases in which $(n', n)$ are even and/or  odd numbers, which are labelled from $1$ to $4$.
\begin{itemize}
\item Case $1$ ( $n'=$ even and $n=$ even): when $n'$ is an even integer the adjoint particle is created inbetween two different quasiparticle pairs, thus we have a sequence of $n/2$ quasiparticle pairs which are all forced to have their quantum numbers equal to $\lambda_{0}\equiv 0$. 
The average over the set of quantum numbers can be written as:
\be
\begin{array}{lcl}
\label{eq:51bis}
 P_{1}(\{k_{\nu}\}) &=& {\displaystyle\sum_{\lambda'_{1}, \lambda''_{1}}\sum_{\lambda'_{2}, \lambda''_{2}}\dots\sum_{\lambda'_{M}, \lambda''_{M}}\prod_{\nu=1}^{M}}\,P^{\lambda'_{\nu}\lambda''_{\nu}}(k_{\nu})\\
&& \times\left[\delta_{0,\lambda_{p+n'+1}}  \delta_{\lambda_{p+n'+1},\lambda_{p+n'+2}}\dots\delta_{\lambda_{p+n'+n-1},\lambda_{p+n'+n}}\right].
\end{array}
\ee
It is worth noticing that the labels in the Kronecker deltas in the previous equation refer to the quasiparticles, while the labels in the summations and in the integrals refer to quasiparticle pairs. In order not to mix up these two different notations we recognize that in this case $p+n'$ is an even number, therefore particle with label $p+n'+1$ belongs to the $\frac{p+n'}{2}+1$ pair, and so on. From this pair to the right we have a sequence of $n/2$ quasiparticle pairs all with quantum numbers $\lambda=0$. In addition the probability distribution factorizes into single-pair probabilities.
Let us remember that the average is taken over all possible initial conditions, therefore $k_{\nu}$ specifies the value of the momentum of the particle at time $t=0+\epsilon$, where $\epsilon$ is a small positive quantity.  Roughly speaking $k_{\nu}$ is the value of the momentum of a particle right after time $t=0$ and before the first scattering process.  Equation (\ref{eq:51bis}) can be easily evaluated to give:
\be
\label{eq:52bis}
\begin{array}{lcl}
 P_{1}(\{k_{\nu}\}) &=& P(k_{1})\dots P\left( k_{\frac{p+n'}{2}}\right)\underbrace{P^{00}\left( k_{\frac{p+n'}{2}+1}\right)\dots P^{00}\left (k_{\frac{p+n'}{2}+\frac{n}{2}}\right)}_{\frac{\left|n\right|}{2}{\rm times}}\\
& & \times P\left(k_{\frac{p+n'}{2}+\frac{n}{2}+1}\right)\dots P(k_{M}),
\end{array}
\ee
where $P(k)={\displaystyle\sum_{a, b=0,\pm 1}}P^{ab}(k)$.
\begin{figure}[t]
   \begin{center}
   \includegraphics[width=4cm]{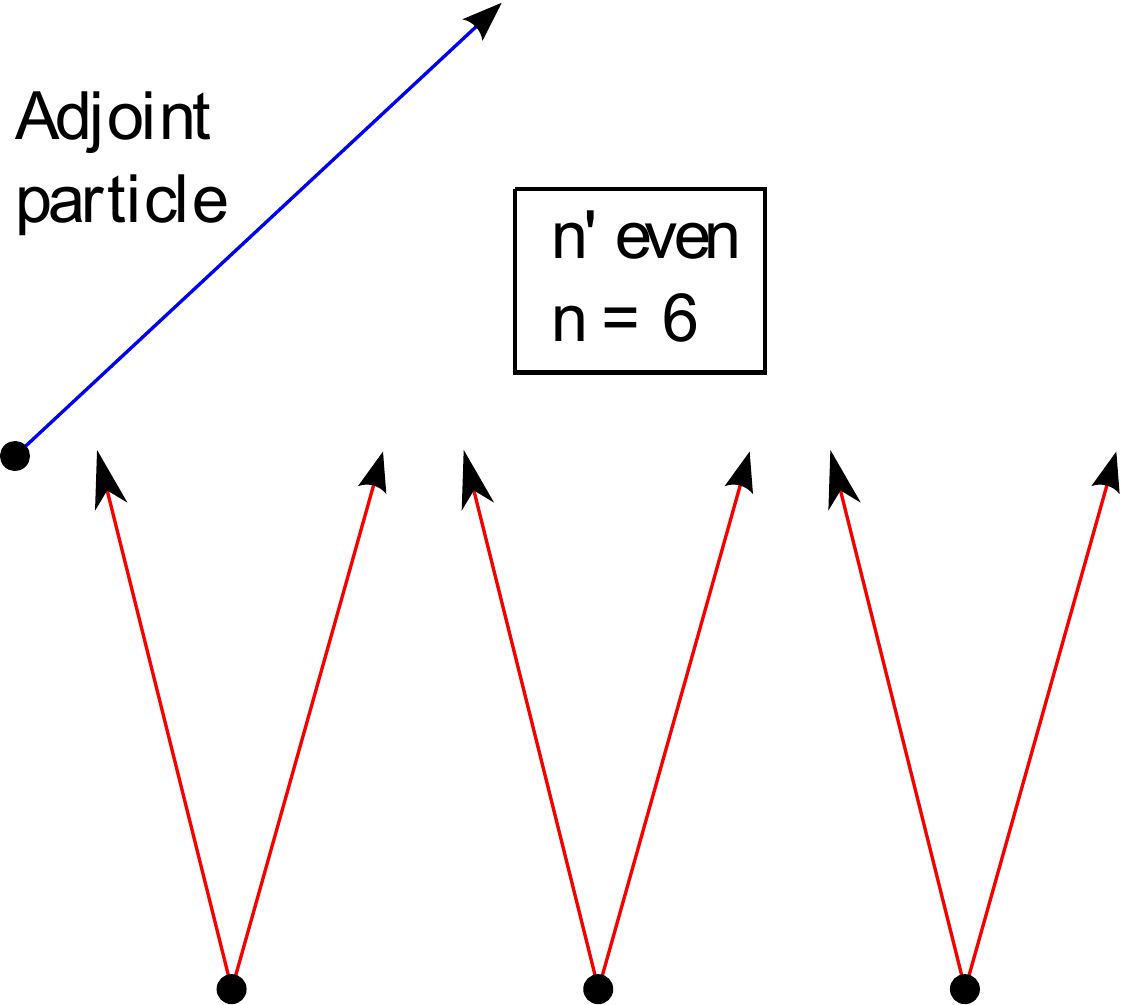}
   \caption{Computation of $P_{1}(\{k_{\nu} \})$: an example with $n=6$. Red trajectories represent those particles whose quantum number must be equal to $\lambda_{0}=0$. }
\label{fig:5}
\end{center}
\end{figure}

\item Case $2$ ($n'=$ even and $n=$ odd): also in this case the adjoint particle is created inbetween two different pairs, but when $n$ is odd the result of equation (\ref{eq:51bis}) becomes:
\begin{equation}
\label{eq:53bis}
\begin{array}{lcl}
P_{2}(\{ k_{\nu}\}) &  = & P(k_{1})\dots P\left ( k_{\frac{p+n'}{2}}\right)\underbrace{P^{00}\left( k_{\frac{p+n'}{2}+1}\right)\dots P^{00}\left (k_{\frac{p+n'}{2}+\frac{n-1}{2}}\right)}_{\frac{|n|-1}{2} {\rm times}} \\
& \times & {\displaystyle\sum_{\lambda=0,\pm 1}} P^{0\lambda}\left( k_{\frac{p+n'}{2}+\frac{n-1}{2}+1}\right) P\left ( k_{\frac{p+n'}{2}+\frac{n-1}{2}+2}\right )\dots P(k_{M}).
\end{array}
\end{equation}
In practice the last particle of the sequence of $\lambda_{\nu}=0$ belongs to a pair whose right-side partner can carry all possible quantum numbers.

\item Case $3$ ($n'=$ odd and $n=$ even): let us consider the case in which $n'$ is an odd integer. In this case the adjoint particle is created within a quasiparticle pair, and the average over quantum numbers gives ($n\ne 0$):
\begin{equation}
\label{eq:54bis}
\begin{array}{lcl}
 P_{3}(\{ k_{\nu}\}) & = & P(k_{1})\dots P\left ( k_{\frac{p+n'-1}{2}}\right)
{\displaystyle\sum_{\lambda=0,\pm 1}} P^{\lambda 0}\left( k_{\frac{p+n'-1}{2}+1}\right)\\
& \times & \underbrace{P^{00}\left( k_{\frac{p+n'-1}{2}+2}\right)\dots P^{00}\left (k_{\frac{p+n'-1}{2}+\frac{n}{2}}\right)}_{\frac{|n|}{2}-1 {\rm times}}\\
& \times & {\displaystyle\sum_{\lambda=0,\pm 1}} P^{0\lambda}\left( k_{\frac{p+n'-1}{2}+\frac{n}{2}+1}\right) P\left ( k_{\frac{p+n'-1}{2}+\frac{n}{2}+2}\right )\dots P(k_{M}).
\end{array}
\end{equation}
The case $n =0$ simply gives a sequence of $M$ different $P(k_{\nu})$'s.
When $n'$ is odd and $n$ even in the sequence of quantum numbers with $\lambda_{\nu}=0$ the first and the last one belong to pairs whose partner has no constraint on its quantum number $\lambda$. This is the reason for the factors ${\displaystyle\sum_{\lambda=0,\pm 1}} P^{0\lambda}(k)$ and ${\displaystyle\sum_{\lambda=0,\pm 1}} P^{\lambda 0}(k)$ in the previous expression. 

\item Case $4$ ($n'=$ odd and $n=$ odd): in this case the average over the set of quantum numbers gives:
\begin{equation}
\label{eq:55bis}
\begin{array}{lcl}
P_{4}(\{ k_{\nu}\}) &  = & P(k_{1})\dots P\left ( k_{\frac{p+n'-1}{2}}\right)
{\displaystyle\sum_{\lambda=0,\pm 1}} P^{\lambda 0}\left( k_{\frac{p+n'-1}{2}+1}\right)\\
& \times & \underbrace{P^{00}\left( k_{\frac{p+n'-1}{2}+2}\right)\dots P^{00}\left (k_{\frac{p+n'-1}{2}+\frac{n-1}{2}+1}\right)}_{\frac{|n|-1}{2} {\rm times}}\\
& \times & P\left ( k_{\frac{p+n'-1}{2}+\frac{n-1}{2}+2}\right )\dots P(k_{M}).
\end{array}
\end{equation}
\end{itemize}
We are now ready to write down the global expression for the quantum number average:
\begin{equation}
\label{eq:56bis}
\begin{array}{l}
\fl {\displaystyle\sum_{\lambda'_{1}, \lambda''_{1}}\sum_{\lambda'_{2}, \lambda''_{2}}\dots\sum_{\lambda'_{M}, \lambda''_{M}}\prod_{\nu=1}^{M}}\,P^{\lambda'_{\nu}\lambda''_{\nu}}(k_{\nu})\, \delta_{0,\lambda_{p+n'+1}} \delta_{\lambda_{p+n'+1},\lambda_{p+n'+2}}\dots\delta_{\lambda_{p+n'+n-1},\lambda_{p+n'+n}}\\
={\displaystyle\frac{1+(-1)^{n'}}{2}}\left[{\displaystyle\frac{1+(-1)^{n}}{2}} P_{1}(\{k_{\nu}\})+ {\displaystyle\frac{1-(-1)^{n}}{2}} P_{2}(\{k_{\nu}\})\right]\\
+{\displaystyle\frac{1-(-1)^{n'}}{2}}\left[{\displaystyle\frac{1+(-1)^{n}}{2}} P_{3}(\{k_{\nu}\})+ {\displaystyle\frac{1-(-1)^{n}}{2}} P_{4}(\{k_{\nu}\})\right].
\end{array}
\end{equation}
This expression is a function of $(n, n')$ and the set of quantum momenta of a particular quasiparticle configuration.  It is worth noticing that $n$ and $n'$ enter the expression for $P_{i}$, $i=1,\dots,4$ in different ways. While $n$ tells us how long the sequence is, $n'$ gives us information about the starting point of the sequence and can easily be absorbed into the definition of $p$. As we shall see, for the actual computation of the relaxation function $R$ the starting position of the sequence is not important, what matters is only the length of the sequence itself. 
\begin{figure}[t]
   \begin{center}
   \includegraphics[width=4cm]{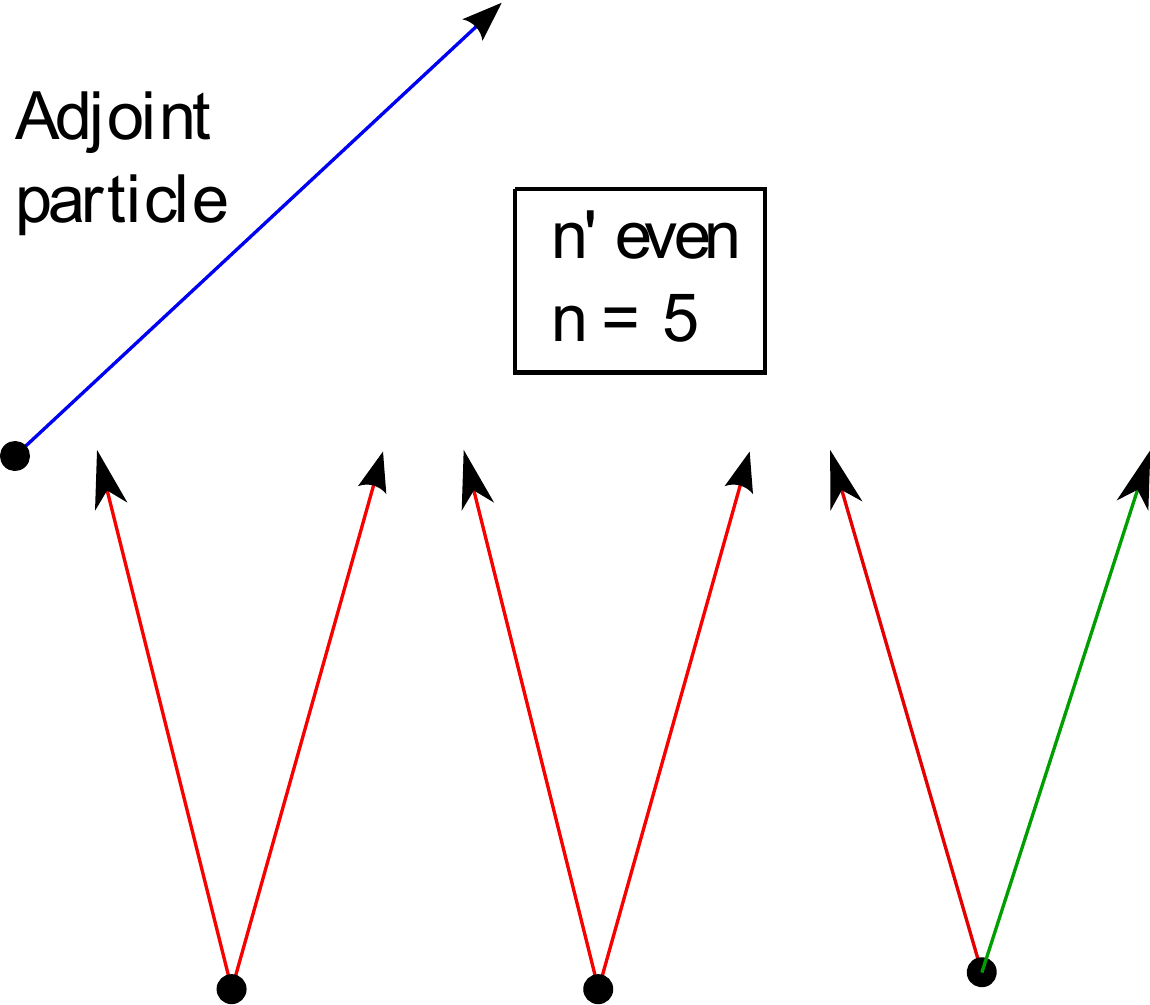}
   \caption{Computation of $P_{2}(\{k_{\nu} \})$: an example with $n=5$.}
\label{fig:5bis}
\end{center}
\end{figure}
\subsection{The semi-classical average in action}
\begin{figure}[t]
\begin{center}
   \includegraphics[width=12cm]{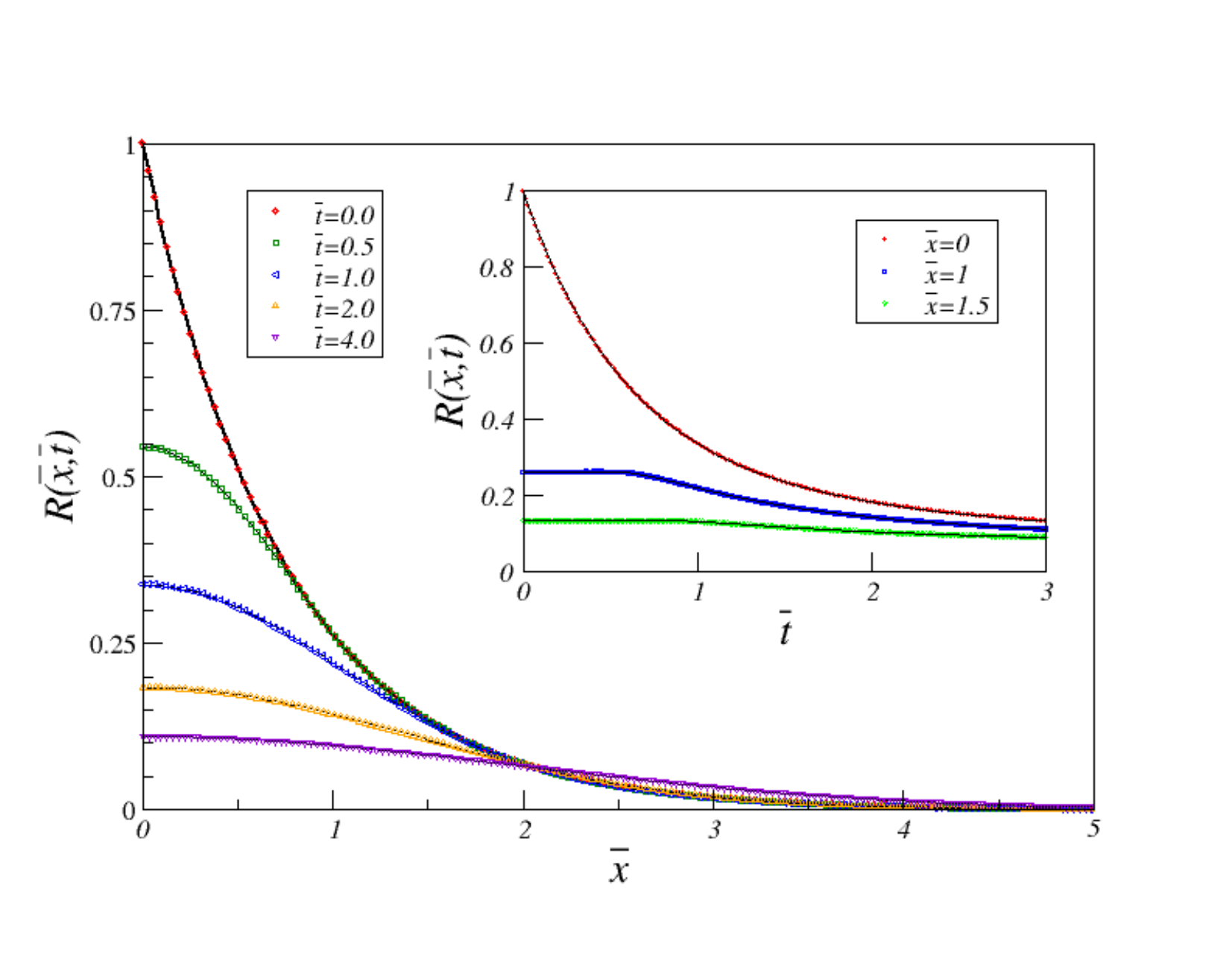}
   \caption{The relaxation function $R(\bar{x}, \bar{t})$ for $f_{k}= A\,{\rm e}^{-\beta k^{2}}$ (where $A=0.1$ and $\beta=0.8$ have been used in this case) and $P^{a,b}= 1/9$ for any choice of quantum numbers. Space and time are measured in unit of $\xi$ and $\tau$ respectively. Data points are numerical results and the solid line is the analytical prediction for the two-point function of a thermal initial state, see Eq. (\ref{eq:74tris}).}
\label{fig:numerical}
\end{center}
\end{figure}

We must now compute the average of the previous quantity over all possible initial positions and momenta. Namely we must take the following average:
\begin{equation}
\label{eq:57bis}
\fl\qquad{\displaystyle\int_{-L/2}^{L/2}\frac{dx_1}{L}\dots\frac{dx_M}{L}\sum_{ k_{1}>0}\dots\sum_{ k_{M}>0}}\,\,
{\displaystyle\sum_{n=-\infty}^{+\infty}}(-1)^{n}\delta_{n, \sum_{\nu_{2}}}\left [\dots\right]{\displaystyle\sum_{n'=-\infty}^{+\infty}}\delta_{n', \sum_{\nu_{1}}}\left [\dots\right],
\end{equation}
where the quantity in the square brackets is given by equation (\ref{eq:56bis}). We can split this average into four pieces, let us call them $A, B, C$ and $D$, one for each $P_{i}(\{ k_{\nu}\})$ respectively, where $i=1,\dots,4$. We shall do the computation of the term which contains $P_{1}(\{k_{\nu} \})$ in some detail; the other cases are straightforward modifications.\\
Let us compute explicity $A$, the first addend of the right side of equation (\ref{eq:56bis}), where the first summation over $n'$ gets immediately canceled out  by the corresponding Kronecker delta because $n'$ itself enters the function $P_{1}(\{k_{\nu} \})$ trivially, therefore we have:
\begin{equation}
\label{eq:59bis}
\begin{array}{lcl}
A&=&{\displaystyle\int_{-L/2}^{L/2}\frac{dx_1}{L}\dots\frac{dx_M}{L}\sum_{ k_{1}>0}\dots\sum_{ k_{M}>0}}\,\,
{\displaystyle\sum_{n=-\infty}^{+\infty}}(-1)^{n}\delta_{n, \sum_{\nu_{2}}}\left [\dots\right]\\
&& \times \left( {\displaystyle\frac{1+(-1)^{\sum_{\nu_{1}}[\dots]}}{2}} {\displaystyle\frac{1+(-1)^{n}}{2}} P_{1}(\{k_{\nu}\}) \right),
\end{array}
\end{equation}
\begin{figure}[t]
   \begin{center}
   \includegraphics[width=4cm]{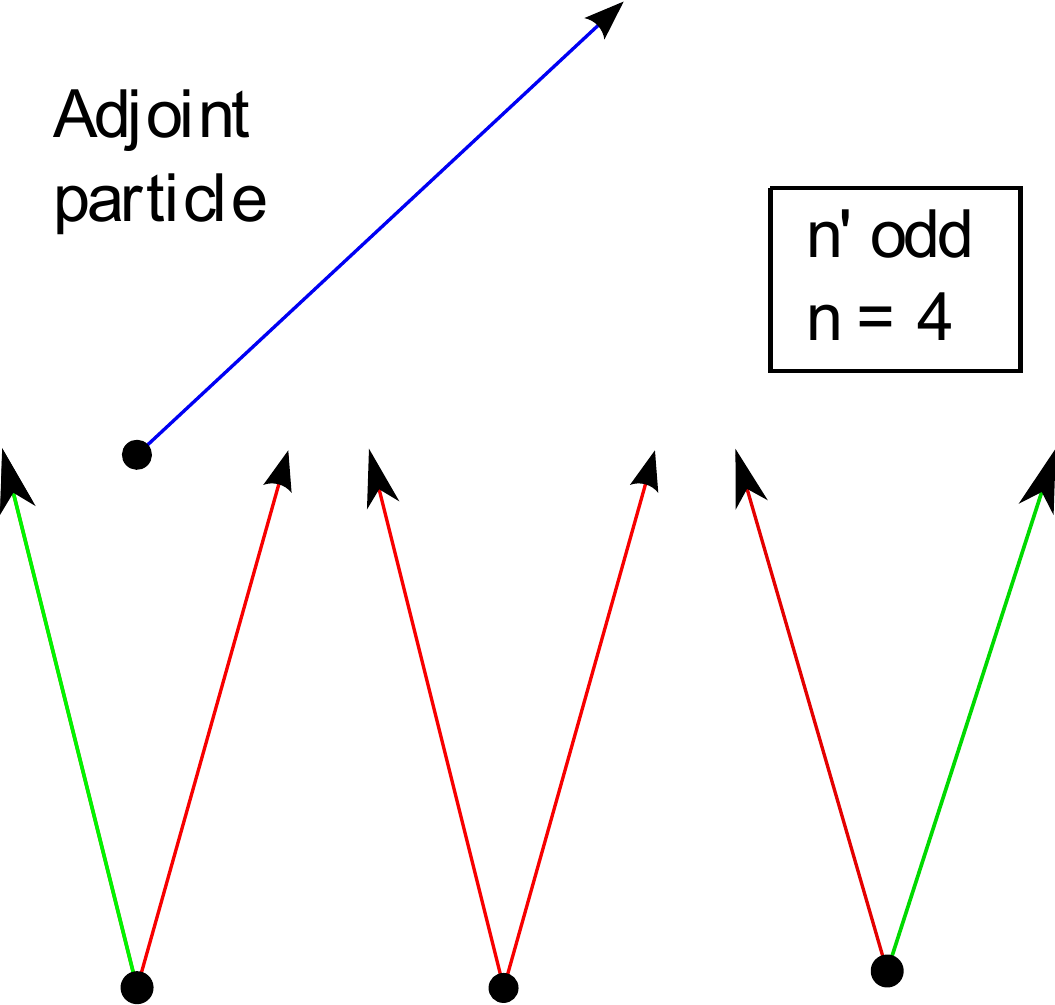}
   \caption{Computation of $P_{3}(\{k_{\nu} \})$: an example with $n'$ odd and $n=4$.}
\label{fig:5tris}
\end{center}
\end{figure}
where $\sum_{\nu_{1}}[\dots]$ is specified by (\ref{eq:43bis}). $A$ can itself be splitted into two pieces, say $A_{1}$ and  $A_{2}$, with $A=A_1+A_2$ which correspond to the two contributions to the integration of the terms into the bracket, namely:
\begin{equation}
\label{eq:60bis1}
\begin{array}{lcl}
A_1 &=&{\displaystyle\int_{-L/2}^{L/2}\frac{dx_1}{L}\dots\frac{dx_M}{L}\sum_{ k_{1}>0}\dots\sum_{ k_{M}>0}}\,\,
{\displaystyle\sum_{n=-\infty}^{+\infty}}(-1)^{n}\delta_{n, \sum_{\nu_{2}}}\left [\dots\right]\\
&& \times \left( {\displaystyle\frac{1+(-1)^{n}}{4}} P_{1}(\{k_{\nu}\}) \right),
\end{array}
\end{equation}
and
\begin{equation}
\label{eq:60bis2}
\begin{array}{lcl}
A_2 &=&{\displaystyle\int_{-L/2}^{L/2}\frac{dx_1}{L}\dots\frac{dx_M}{L}\sum_{ k_{1}>0}\dots\sum_{ k_{M}>0}}\,\,
{\displaystyle\sum_{n=-\infty}^{+\infty}}(-1)^{n}\delta_{n, \sum_{\nu_{2}}}\left [\dots\right]\\
&& \times \left( {\displaystyle\frac{(-1)^{\sum_{\nu_{1}}[\dots]}}{2}} {\displaystyle\frac{1+(-1)^{n}}{2}} P_{1}(\{k_{\nu}\}) \right).
\end{array}
\end{equation}
Let us start by computing the contribution of $A_1$. By making use of the following integral representation of the Kronecker delta
\be
\label{int_rep}
\delta_{n,\sum_{\nu}[\dots]} = \frac{1}{2\pi}{\displaystyle\int_{-\pi}^{\pi}}d\phi\,{\rm e}^{i\phi(n-\sum_{\nu}[\dots])},
\ee
 we can write $A_1$ as:
\begin{equation}
\label{eq:61bis}
\fl\quad\int_{-\pi}^{\pi}\frac{d\phi}{2\pi}{\displaystyle\sum_{n=-\infty}^{+\infty}}(-1)^{n}{\rm e}^{i n\phi}{\displaystyle\frac{1+(-1)^{n}}{2}}{\displaystyle\sum_{ k_{1}>0}\dots\sum_{ k_{M}>0}}\,\,{\displaystyle\int_{-L/2}^{L/2}\frac{dx_1}{L}\dots\frac{dx_M}{L}}\,{\rm e}^{-i\phi\sum_{\nu_{2}}[\dots]}\,\frac{P_{1}(\{k_{\nu}\})}{2}.
\end{equation}
Now we first compute the spatial integration in equation \ref{eq:61bis}, then the integral over the momenta and finally the sum over $n$.
\begin{figure}[t]
   \begin{center}
   \includegraphics[width=4cm]{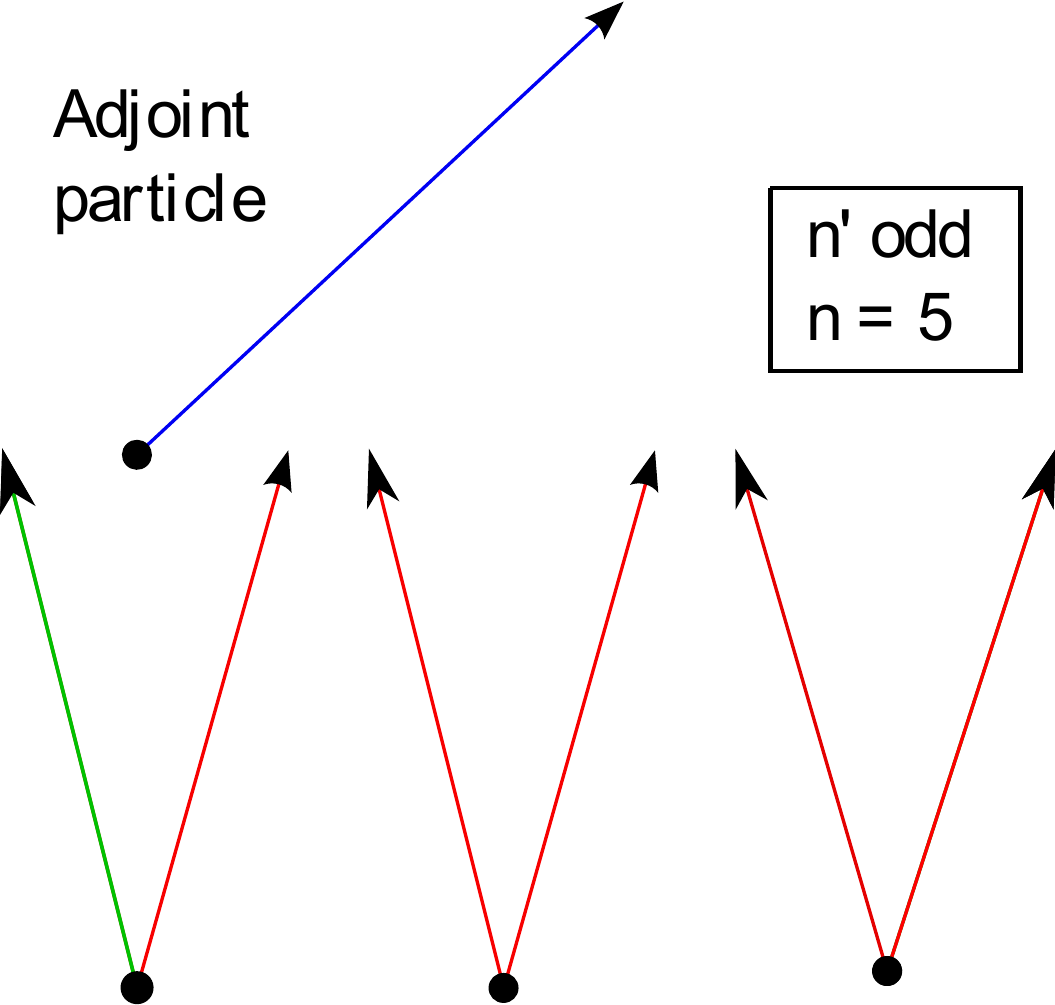}
   \caption{Computation of $P_{4}(\{k_{\nu} \})$: an example with $n'$ odd and $n=5$.}
\label{fig:5quatris}
\end{center}
\end{figure}
\paragraph{The spatial integration.}
Now we want to compute \emph{the spatial integral}, remembering that the sum $\sum_{\nu_{2}}[\dots]$ is specified in equation (\ref{eq:45bis}). We begin by considering $x>0$, $t_{2}>t_{1}$ and $x<v_{max}(t_2-t_1)$, where $v_{max}$ is the maximal velocity of the excitations of the model (in the appendix B we will show the general result for arbitrary times and distances). We will refer to this situation as the within-the-light-cone case.
The spatial integral factorizes in a straightforward way, thus we can write:
\begin{equation}
\label{eq:63bis}
\begin{array}{l}
\fl\quad{\displaystyle\int_{-L/2}^{L/2}\frac{dx_1}{L}\dots\frac{dx_M}{L}}\,{\rm e}^{-i\phi\sum_{\nu_{2}}[\dots]}={\displaystyle\prod_{\nu=1}^{M}}\left\{ 1-\frac{2v_{k_{\nu}}t}{L}(1-\cos\phi)\Theta[v_{k_{\nu}}t-x]\right.\\
\\
\fl\quad\left.-\frac{2ix\sin\phi}{L}\Theta[v_{k_{\nu}}t-x]-\frac{2x}{L} \left(i\sin\phi+2\sin^{2}(\phi/2)\right)\Theta[x-v_{k_{\nu}}t] \Theta[v_{k_{\nu}}T-x]\right.\\
\\
\fl\quad-\left.\left(\frac{v_{k_{\nu}}T}{L}(1+\cos(2\phi)-i\sin(2\phi)-\cos\phi+i\sin\phi)+ \frac{x}{L}(i\sin(2\phi)+2\sin^{2}\phi) \right) \Theta[x-v_{k_{\nu}}T]\right\},
\end{array}
\end{equation}
where $v_{k_{\nu}}$ is the particle velocity defined by $v_k\equiv\frac{\partial\epsilon(k)}{\partial k}$, $t\equiv t_2-t_1$ and $T\equiv t_2 + t_1$. 
Let us notice the following definition of the probability $P(k)$:
\be
\label{eq.def}
P(k)\equiv{\displaystyle\frac{f_k}{M}},\qquad M=\sum_{k>0}f_k, 
\ee
where $f_k$ is the occupation number of the $k$-mode. Furthermore, the following identities can be useful in taking the thermodynamic limit:
\begin{equation}
\label{eq:67bis}
\fl\quad{\displaystyle\frac{{\displaystyle\sum_{k>0}P(k) v_k}}{L}}\equiv{\displaystyle\frac{{\displaystyle\sum_{k>0}f_k v_k}}{L{\displaystyle\sum_{k>0}f_k}}}\longrightarrow{\displaystyle\frac{{\displaystyle\int_{0}^{\infty}\frac{dk}{2\pi}f_k v_k}}{M}},\qquad\qquad{\displaystyle\frac{{\displaystyle\sum_{k>0}P(k)}}{L}}\equiv{\displaystyle\frac{{\displaystyle\sum_{k>0}f_k}}{L{\displaystyle\sum_{k>0}f_k}}}\longrightarrow{\displaystyle\frac{{\displaystyle\int_{0}^{\infty}\frac{dk}{2\pi}f_k}}{M}}.
\end{equation}
\paragraph{Integration over the momenta.}
Let us now compute the average over {\emph the momenta} of the previous result, taking the limit $L,M \to\infty$ ($M/L\equiv\rho$ fixed). The basic idea behind this limit is that we can replace $M$ by the average particle number, $M \to \rho L$ without changing the final result. Namely we have:
\begin{equation}
\label{eq:64bis}
{\displaystyle\lim_{M,L\to\infty}}{\displaystyle\sum_{ k_{1}>0}\dots\sum_{ k_{M}>0}}P_{1}(\{k_{\nu}\}){\displaystyle\prod_{\nu=1}^{M}}\left\{\dots\right\},
\end{equation}
where $P_{1}(\{k_{\nu}\})$ is given by (\ref{eq:52bis}) and the $\{\dots\}$ are the right-hand terms of equation (\ref{eq:63bis}). This expression can be simply evaluated to yield:
\begin{equation}
\label{eq:69bis}
\begin{array}{l}
\fl{\displaystyle\lim_{M,L\to\infty}}\,{\displaystyle\sum_{ \{k_{\nu}>0\}}P_{1}(\{k_{\nu}\})}{\displaystyle\prod_{\nu=1}^{M}}\left( \dots\right) \\
\fl=\exp{\left\{-t(1-\cos\phi){\displaystyle\int_{0}^{\infty}\frac{dk}{\pi}f_k\, v_{k}}\Theta[v_{k}t-x]-2x\sin^{2}(\phi/2){\displaystyle\int_{0}^{\infty}\frac{dk}{\pi}f_k}\,\Theta[x-v_{k}t]\Theta[v_{k}T-x]\right\}}\\
\fl\quad\times\exp{\left\{-T(1+\cos(2\phi)-i\sin(2\phi)-2\cos\phi+2i\sin\phi){\displaystyle\int_{0}^{\infty}\frac{dk}{2\pi}f_k}\,\Theta[x-v_{k}T]\right\}}\\
\fl\quad\times\exp{\left\{-i x\sin\phi{\displaystyle\int_{0}^{\infty}\frac{dk}{\pi}f_k}\,(\Theta[v_{k}t-x]+\Theta[x-v_{k}t]\Theta[v_{k}T-x])\right\}}\\
\fl\quad\times\exp{\left\{-x(2\sin^{2}\phi+i\sin(2\phi)){\displaystyle\int_{0}^{\infty}\frac{dk}{2\pi}f_k}\,\Theta[x-v_{k}T]\right\}}(P^{00})^{|n|/2},\\ \\
\end{array}
\end{equation}
where $P^{00}=\int_{0}^{\infty}\frac{dk}{2\pi}P^{00}(k)$ and $P^{00}(k)$ is now a probability {\em density}. 
We can now take the limit when $T$ is very large ($x/T \ll v_{max}$), assuming that $T\int_{0}^{\infty}\frac{dk}{2\pi}f_k v_k\,\Theta[x-v_{k}T]\to 0$, so the previous expression simplifies, giving:
\begin{equation}
\label{eq:69tris}
\begin{array}{lcl}
\fl\dots & = &  \exp{\left\{-t(1-\cos\phi){\displaystyle\int_{0}^{\infty}\frac{dk}{\pi}f_k\, v_{k}}\Theta[v_{k}t-x]-2x\sin^{2}(\phi/2){\displaystyle\int_{0}^{\infty}\frac{dk}{\pi}f_k}\,\Theta[x-v_{k}t]\right\}}\\
\fl& \times &  \exp{\left\{-i x\sin\phi{\displaystyle\int_{0}^{\infty}\frac{dk}{\pi}f_k}\right\}}\,(P^{00})^{|n|/2}.
\end{array}
\end{equation}
 Before computing the sum over $n$ it is convenient to obtain the analogy of equation (\ref{eq:69tris}) for the case $A_{2}$, and see which contributions survive in the limit of $T$ very large . For $A_2$ we start with:
\begin{equation}
\label{eq:70bis}
{\displaystyle\sum_{ k_{1}>0}\dots\sum_{ k_{M}>0}}\,{\displaystyle\int_{-L/2}^{L/2}\frac{dx_1}{L}\dots\frac{dx_M}{L}}\,{\rm e}^{-i\phi\sum_{\nu_{2}}[\dots]}\,P_{1}(\{k_{\nu}\})\,(-1)^{\sum_{\nu_{1}}[\dots]}, 
\end{equation}
then by repeating the same steps we end up with:
\begin{equation}
\label{eq:72bis}
\begin{array}{l}
{\displaystyle\lim_{M\to\infty}}\,{\displaystyle\sum_{ k_{1}>0}\dots\sum_{ k_{M}>0}}P_{1}(\{k_{\nu}\}){\displaystyle\prod_{\nu=1}^{M}}\left( \dots\right) \\
=\exp{\left(-(T-t\cos\phi){\displaystyle\int_{0}^{\infty}\frac{dk}{2\pi}f_k v_k}-i x\sin\phi{\displaystyle\int_{0}^{\infty}\frac{dk}{2\pi}f_k\,\Theta[v_k t-x]}\right)}\\
\times\exp{\left(-i t\sin\phi{\displaystyle\int_{0}^{\infty}\frac{dk}{2\pi}f_k v_k\,\Theta[x-v_k t]}\right)}\,(P^{00})^{|n|/2}.
\end{array}
\end{equation}
We immediately see that this contribution is exponentially decreasing as $T$ becomes large, therefore we neglect it in this limit.
\paragraph{Summation over $n$.}
Performing now the summation over $n$, the global contribution from term $A$ to the relaxation function (\ref{eq:49bis}) is given by:
\begin{equation}
\label{eq:73bis}
\begin{array}{l}
\fl A =  {\displaystyle\lim_{\substack{L, M\to\infty}}}\left\langle
{\displaystyle\sum_{n=-\infty}^{+\infty}}(-1)^{n}\delta_{n, \sum_{\nu_{2}}[\dots]}\left( \frac{1+(-1)^{\sum_{\nu_{1}}}}{2} \frac{1+(-1)^{n}}{2} P_{1}(\{k_{\nu}\}) \right)\right\rangle_{\{x_{\nu},k_{\nu}\}}\\
\fl\quad\:\:\, = {\displaystyle\int_{-\pi}^{+\pi}\frac{d\phi}{4\pi}}\,\,{\rm e}^{-t(1-\cos\phi)\int_{0}^{\infty}\frac{dk}{\pi}f_k\, v_{k}\Theta[v_{k}t-x]}\,{\rm e}^{-2x\sin^{2}(\phi/2)\int_{0}^{\infty}\frac{dk}{\pi}f_k\,\Theta[x-v_{k}t]}\\
\fl\quad\:\:\,\times \cos\left(x\sin\phi\int_{0}^{\infty}\frac{dk}{\pi}f_k\right){\displaystyle\frac{1-(P^{00})^{2}}{1-2P^{00}\cos(2\phi)+(P^{00})^{2}}},
\end{array}
\end{equation}
which in the large $T$ limit depends only on $t$. Repeating all the steps we did for the term containing $P_{1}(\{k_{\nu}\})$, we can then compute the contributions from the terms containing $P_{2}(\{k_{\nu}\})$,  $P_{3}(\{k_{\nu}\})$ and $P_{4}(\{k_{\nu}\})$ (namely $B$, $C$, and $D$). Here we list the results of these computations (always in the limit $x/T \ll v_{max}$):
\begin{equation}
\label{eq:74bis}
\begin{array}{l}
\fl B = {\displaystyle\lim_{\substack{L, M\to\infty}}}\left\langle
{\displaystyle\sum_{n=-\infty}^{+\infty}}(-1)^{n}\delta_{n, \sum_{\nu_{2}}[\dots]}\left( \frac{1+(-1)^{\sum_{\nu_{1}}}}{2} \frac{1-(-1)^{n}}{2} P_{2}(\{k_{\nu}\}) \right)\right\rangle_{\{x_{\nu},k_{\nu}\}}\\
\fl\quad\:\:\,\,={\displaystyle\int_{-\pi}^{+\pi}\frac{d\phi}{4\pi}}\, {\rm e}^{-t(1-\cos\phi)\int_{0}^{\infty}\frac{dk}{\pi}f_k\, v_{k}\Theta[v_{k}t-x]}\,\, {\rm e}^{-2x\sin^{2}(\phi/2)\int_{0}^{\infty}\frac{dk}{\pi}f_k\,\Theta[x-v_{k}t]}\\
\fl\quad\:\:\,\,\times\cos\left(x\sin\phi\int_{0}^{\infty}\frac{dk}{\pi}f_k\right){\displaystyle\frac{2\cos\phi(P^{00}-1)}{1-2P^{00}\cos(2\phi)+(P^{00})^{2}}}{\displaystyle\sum_{\lambda}P^{0\lambda}},
\end{array}
\end{equation}
and
\begin{equation}
\label{eq:74bistris}
\begin{array}{l}
\fl C={\displaystyle\lim_{\substack{L, M\to\infty}}}\left\langle
{\displaystyle\sum_{n=-\infty}^{+\infty}}(-1)^{n}\delta_{n, \sum_{\nu_{2}}[\dots]}\left( \frac{1-(-1)^{\sum_{\nu_{1}}}}{2} \frac{1+(-1)^{n}}{2} P_{3}(\{k_{\nu}\}) \right)\right\rangle_{\{x_{\nu},k_{\nu}\}}\\
\fl\quad\:\:\,\,={\displaystyle\int_{-\pi}^{+\pi}\frac{d\phi}{4\pi}}\,{\rm e}^{-t(1-\cos\phi)\int_{0}^{\infty}\frac{dk}{\pi}f_k\, v_{k}\Theta[v_{k}t-x]}\,\, {\rm e}^{-2x\sin^{2}(\phi/2)\int_{0}^{\infty}\frac{dk}{\pi}f_k\,\Theta[x-v_{k}t]}\\
\fl\quad\:\:\,\,\times\cos\left(x\sin\phi\int_{0}^{\infty}\frac{dk}{\pi}f_k\right){\displaystyle\frac{1+2\cos(2\phi)[(\sum_{\lambda}P^{0\lambda})^{2}-P^{00}]-2P^{00}( \sum_{\lambda}P^{0\lambda})^{2}+(P^{00})^{2}}{1-2P^{00}\cos(2\phi)+(P^{00})^{2}}},
\end{array}
\end{equation}
where we have assumed $P^{0\lambda}\equiv P^{\lambda 0}$, and finally
\begin{equation}
\label{eq:74bisquatris}
\begin{array}{l}
\fl D={\displaystyle\lim_{\substack{L, M\to\infty}}}\left\langle
{\displaystyle\sum_{n=-\infty}^{+\infty}}(-1)^{n}\delta_{n, \sum_{\nu_{2}}}\left( \frac{1-(-1)^{\sum_{\nu_{1}}}}{2} \frac{1-(-1)^{n}}{2} P_{4}(\{k_{\nu}\}) \right)\right\rangle_{\{x_{\nu},k_{\nu}\}}\\
\fl\quad\:\:\,\,={\displaystyle\int_{-\pi}^{+\pi}\frac{d\phi}{4\pi}}\,{\rm e}^{-t(1-\cos\phi)\int_{0}^{\infty}\frac{dk}{\pi}f_k\, v_{k}\Theta[v_{k}t-x]}\,\, {\rm e}^{-2x\sin^{2}(\phi/2)\int_{0}^{\infty}\frac{dk}{\pi}f_k\,\Theta[x-v_{k}t]}\\
\fl\quad\:\:\,\,\times\cos\left(x\sin\phi\int_{0}^{\infty}\frac{dk}{\pi}f_k\right){\displaystyle \frac{2\cos\phi(P^{00}-1)}{1-2P^{00}\cos(2\phi)+(P^{00})^{2}}}{\displaystyle\sum_{\lambda}P^{0\lambda}}.
\end{array}
\end{equation}
Plugging all the pieces together,  $A$+$B$+$C$+$D$, we end up with the expression (\ref{eq:res2}) for the relaxation function :\\
\begin{equation}
\label{eq:74tris}
\begin{array}{l}
\fl\qquad\qquad\qquad R(x; t)={\displaystyle\int_{-\pi}^{\pi}\frac{d\phi}{2\pi}}\exp{\left(-t(1-\cos\phi){\displaystyle\int_{0}^{\infty}\frac{dk}{\pi}f_k\, v_{k}\Theta[v_{k}t-x]}\right)}\\
\\
\fl\qquad\qquad\qquad\times\,\exp{\left(-x(1-\cos\phi){\displaystyle\int_{0}^{\infty}\frac{dk}{\pi}f_k\,\Theta[x-v_{k}t]}\right)}\cos\left(x\sin\phi{\displaystyle\int_{0}^{\infty}\frac{dk}{\pi}f_k}\right)\\
\\
\fl\qquad\qquad\qquad\times\,{\displaystyle\frac{1+2\cos(\phi)(P^{00}-1)Q+\cos(2\phi)(Q^{2}-P^{00})-P^{00}Q^{2}}{1-2P^{00}\cos(2\phi)+(P^{00})^{2}}},
\end{array}
\end{equation}
where $Q= \sum_{\lambda=\pm 1,0}P^{0\lambda}$.
It would be useful to have an obvious comparison between this semi-classical result and a direct GGE computation of the same correlator for the case of the O(3) non-linear sigma model, as was done for the transverse field Ising model and the XY chain in transverse field. The new approach to these kinds of problems introduced in \cite{Caux1},\cite{Caux2} and \cite{Tvelik} may be a missing tool in this respect.\\
Formula (\ref{eq:74tris}) can be tested in several ways, for instance by choosing a thermally-populated initial state, with $f_{k}\propto {\rm e}^{-\beta k^{2}}$, and the same probability for each quantum number  to appear. This check can be done both analytically and numerically. In the former case, starting from expression (\ref{eq:74tris}) and plugging in a thermal distribution for $f_{k}$ and the quantum numbers, we find the same universal analytical result of Reference \cite{Rapp1}, whereas in the latter case  we have performed a numerical average in the same spirit as that in Reference \cite{Damle}. In practice we randomly generate semi-classical configurations in order to compute the average in equation (\ref{eq:49bis}), starting with a system size of $L= 400\times\xi$ and imposing fixed boundary conditions. The density in these units is unity and so the initial state is populated by $400$ particles with their initial positions drawn from a uniform ensemble. The system size is large enough that finite-size effects are negligible for our present purposes. We assign to each particle a velocity from the classical thermal ensemble, and we do the average over the spin values analytically. With this protocol in hand, calculating $R(r,t)$ reduces to some simple bookkeeping that keeps track of the two integers $n$ and $n'$ for a given configuration $C$. We have implemented the numerical average by averaging over $10^{6}$ configurations drawn from the appropriate distribution. The statistical error results to be of order $10^{-4}$, while the results are shown in figure (\ref{fig:numerical}), indicating a perfect agreement between the analitycal and numerical prediction. 

\section{The dynamics of the transverse field Ising model}  %
\label{sec:Ising}
\subsection{The paramagnetic phase}  
\label{sec:subIsing1}
\begin{figure}[t]
\begin{center}
\includegraphics[width=7.5cm]{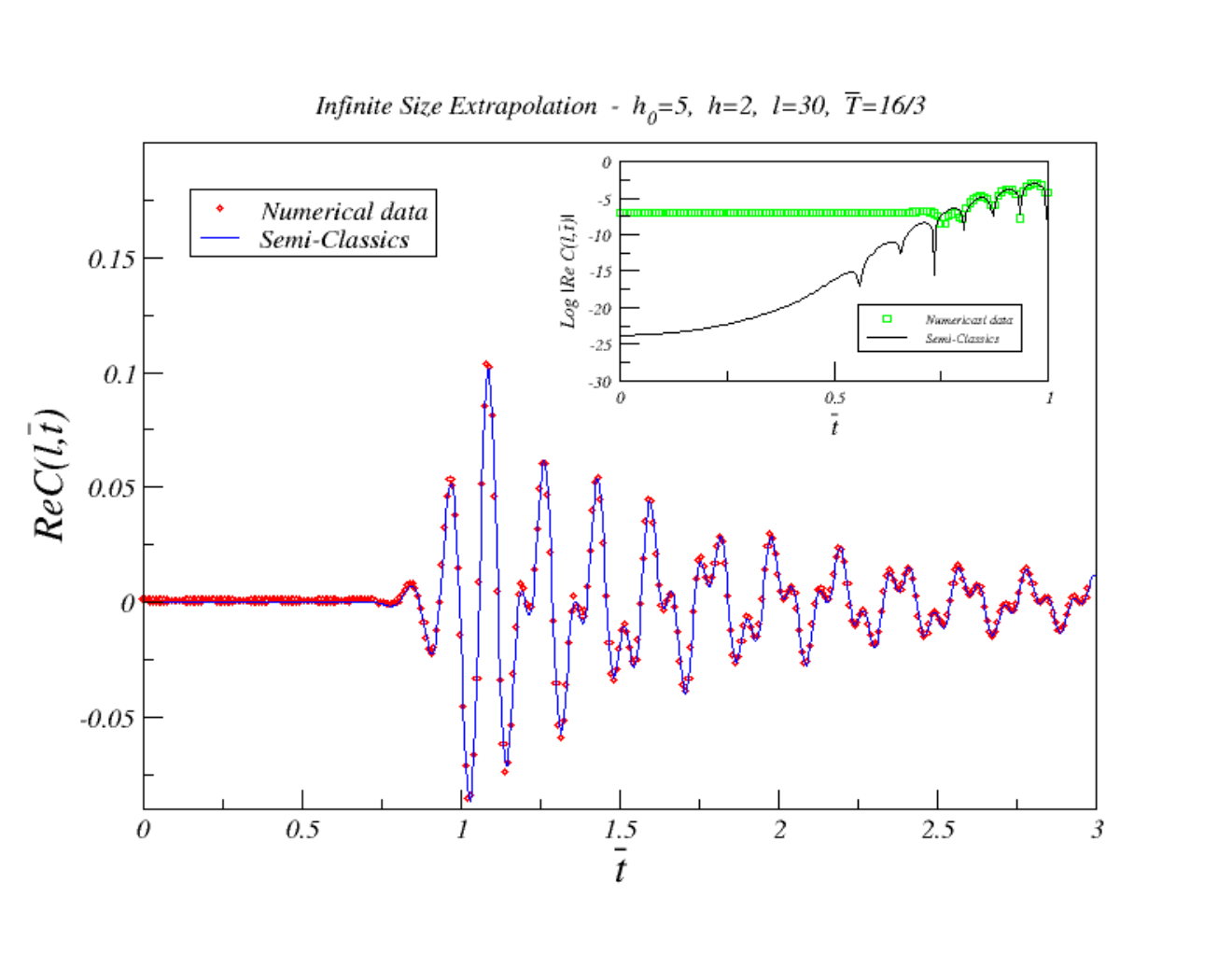}
\includegraphics[width=7.5cm]{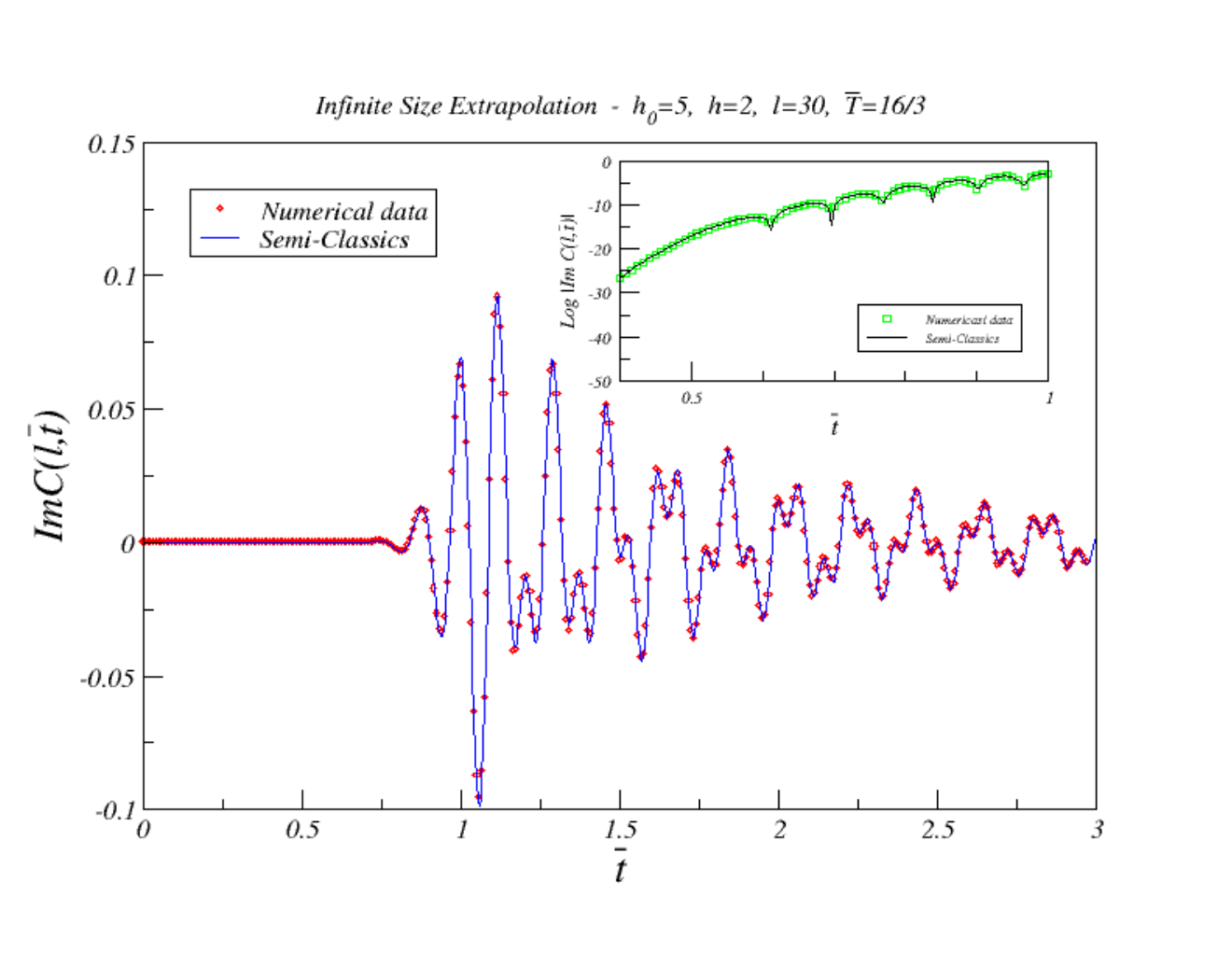}
\end{center}
\caption{The real and imaginary parts of the non-equal-time two point function after a quantum quench in the disordered phase, from $h_0 = 5$ to $h=2$. By definition $\bar{t}=t/t_f$ and $\bar{T}=T/t_f$. Data points represent the numerical data extrapolated in the thermodynamic limit ($L \to \infty$), while the solid line is given by equation (\ref{eq:75}). If we look at the inset in the plot of the real part, we notice that the semi-classical formula does not capture well the behaviour of the correlator outside of the light-cone (the same plot for the imaginary part shows a better agreement, even if at very small values of $\bar{t}$ numerical errors do not allow us to compare analytical result with the numerical ones). }
\label{Fig:numerics1}
\end{figure}

\begin{figure}[t]
\begin{center}
\includegraphics[width=7.5cm]{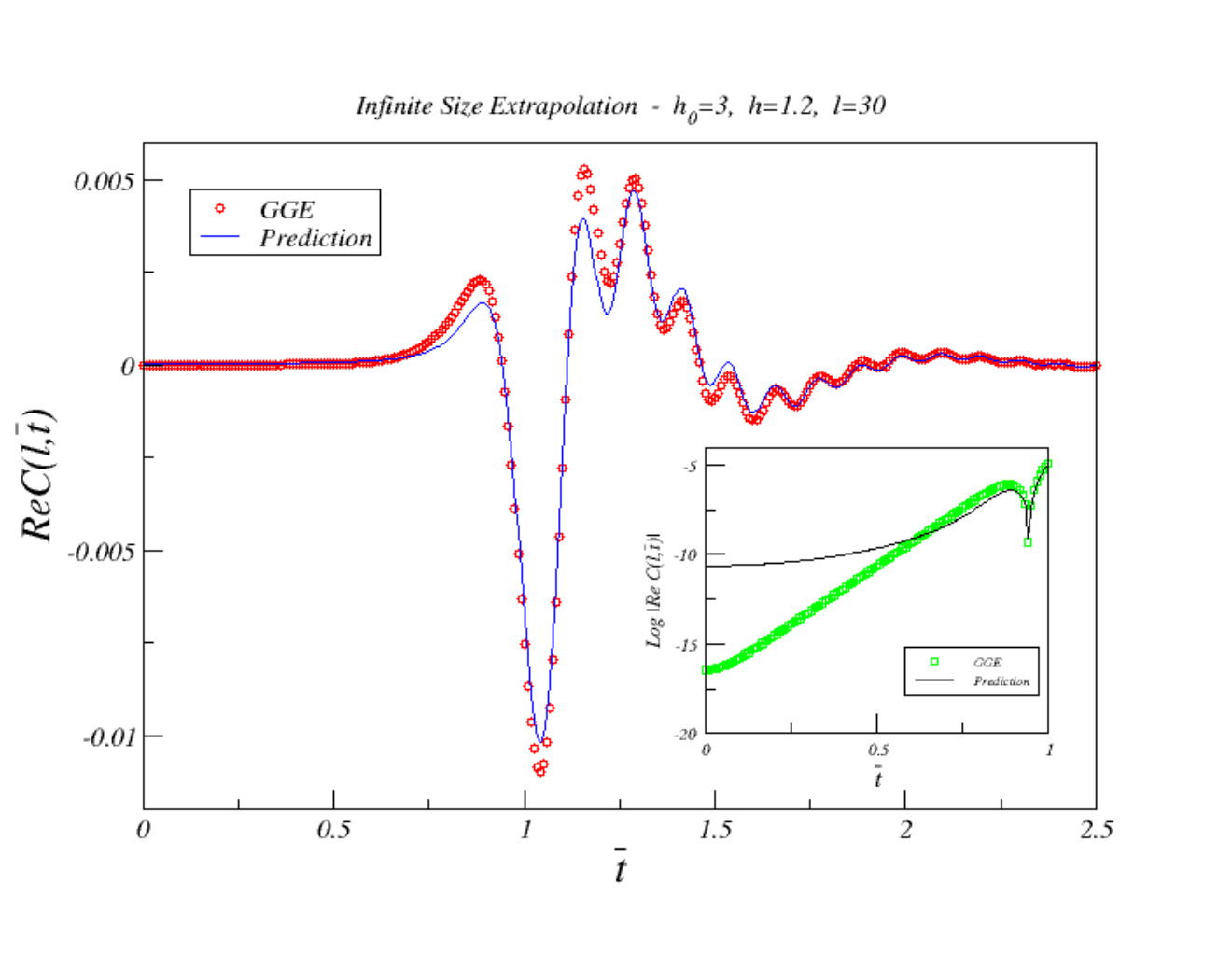}
\includegraphics[width=7.5cm]{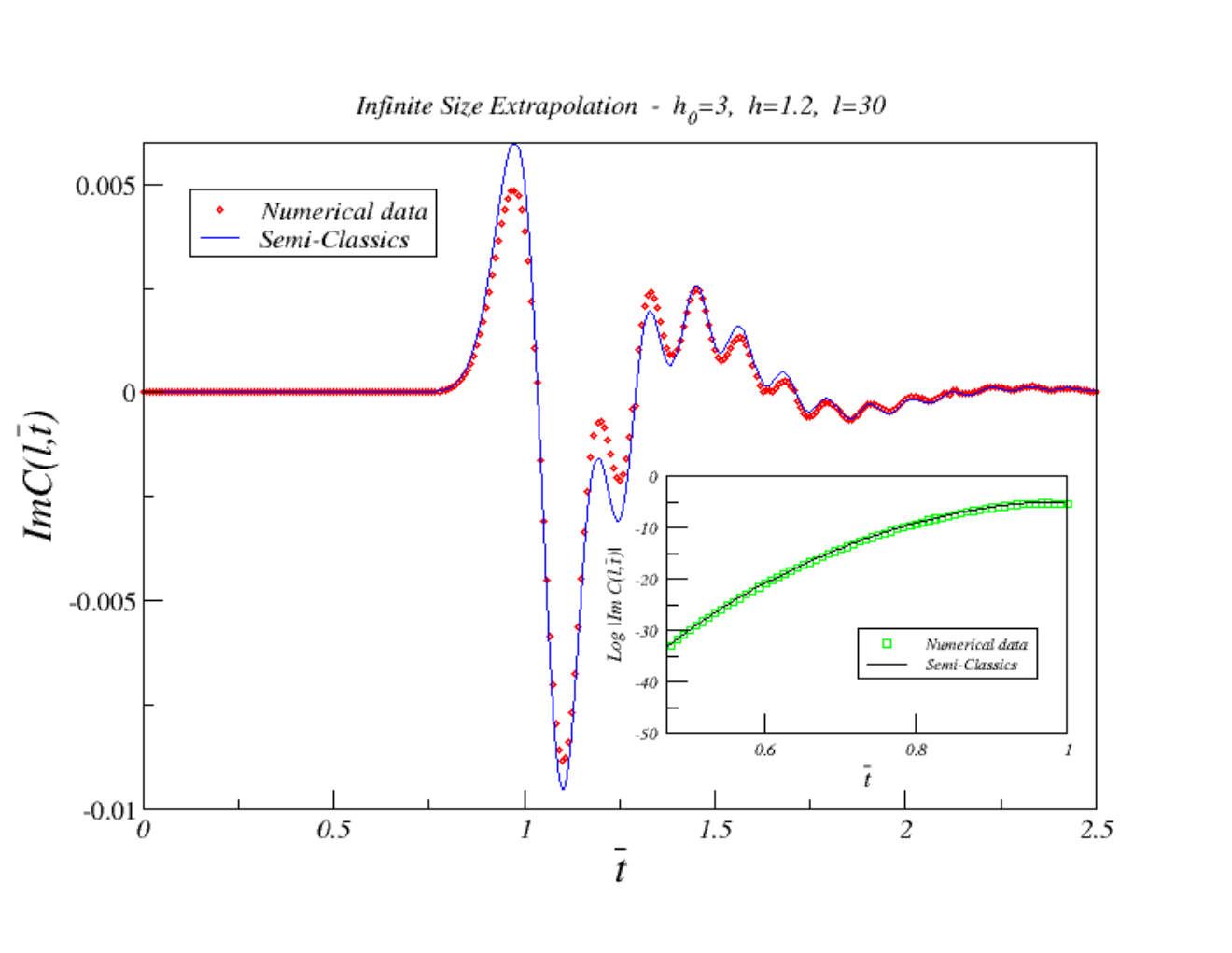}
\end{center}
\caption{The real and imaginary parts of the non-equal-time two point function after a quantum quench in the disordered phase, from $h_0 = 3$ to $h=1.2$. By definition $\bar{t}=t/t_f$ and $\bar{T}/t_f\gg 1$. Data points represent the numerical data extrapolated in the thermodynamic limit ($L \to \infty$), while the solid line is given by equation (\ref{eq:75}). The same considerations made for (\ref{Fig:numerics1}) hold true here. In addition we notice that as this quench is closer to the critical point, the agreement is only qualitative at small $\bar{t}$, despite the substitution $f_k \to -1/2 \log|\cos(\Delta_k)|$. To improve the agreement we should correct the expression for the pre-factor of equation (\ref{eq:75}), following the exact results of reference \cite{Evangelisti}.}
\label{Fig:numerics2}
\end{figure}

Another important check of formula (\ref{eq:74tris}) is represented by the transverse field Ising chain limit (TFIC). Briefly, we now focus on the dynamics after a quantum quench in the TFIC, whose Hamiltonian reads:
\begin{equation}
\label{eq:Ising_ham}
\hat{H}=-J\left({\displaystyle\sum_{i=1}^{L-1}\sigma_{i}^{x}\sigma_{i+1}^{x}}-h{\displaystyle\sum_{i=1}^{L}\sigma_{i}^{z}}\right),
\end{equation}
where $\sigma_{i}^{\alpha}$ are the Pauli matrices at site $i$, $J>0$ is the energy scale, $h$ the transverse magnetic field, and we impose free boundary conditions (for the moment we assume the system to be large but {\em finite}).
The goal is to determine the dynamical order-parameter two-point function:
\be
\label{eq:ising_cor}
C^{\rm Ising}(x, T, t)=\langle \psi |\sigma_{x_{2}}^{x}(t_{2})\sigma_{x_{1}}^{x}(t_{1})| \psi\rangle,
\ee
where $x=x_2-x_1$, $T=t_1+t_2$ and $t=t_2-t_1$. The action of the operator $\sigma_{i}^{x}(t)$ on the paramagnetic ground state is akin to that of the operator $\tilde{n}^{z}$ in the O(3) non-linear sigma model, that is, it either creates an excitation or it destroys one already present. For this reason the semi-classical approach developed in the previous section is easily generalizable.
For this model the $K$ matrix in the definition of $|\psi\rangle$ is known and equal to $K(k)=-\frac{i}{2}\tan\left[\frac{\theta_k-\theta_{k'}}{2}\right]$, where the $\theta$'s are the Bogoliubov angles defined in (\ref{eq:77bis}) before and after the quench respectively. 
In order to obtain (\ref{eq:ising_cor}), without rederiving everything from the beginning, one should formally replace $P^{a,b}=1/(q-1)^{2}$ and then take the limit $q \to 2$. 
Otherwise we can start again from (\ref{eq:56bis}), and take into account that for the TFIC excitations do not have internal quantum numbers (this is true both in the ordered and disordered phases). This feature greatly simplifies the algebra of the derivation, and we end up with the following general expression for the correlator of equation (\ref{eq:ising_cor}), valid in the thermodynamic limit:\\
\begin{equation}
\label{eq:75}
\begin{array}{l}
 \fl C^{\rm Ising}(x, T, t) = C^{\rm Ising}_{T=0}(x, t)\,\exp{\left(-2t{\displaystyle\int_{0}^{\pi}\frac{dk}{\pi}f_k v_k\Theta[v_k t-x]}\right)}\\
\fl\qquad\qquad \times  \exp{\left(-2T{\displaystyle\int_{0}^{\pi}\frac{dk}{\pi}f_k v_k \Theta[x-v_k T]}\right)} \exp{\left(-2x{\displaystyle\int_{0}^{\pi}\frac{dk}{\pi}f_k\Theta[x-v_k t]\Theta[v_k T-x]}\right)},\\
\end{array}
\end{equation}
where $C^{\rm Ising}_{T=0}(x, t)$ is given by (neglecting multi-particle terms in the spectral density \cite{SachdevYoung}):
\begin{equation}
\label{eq:76bis}
 C^{\rm Ising}_{T=0}(x, t)  \propto {\displaystyle\int_{-\pi}^{\pi}\frac{dk}{2\pi}\,\frac{{\rm e}^{-i\epsilon(k)t+ikx}}{\epsilon(k)}},
\end{equation}
where again as long as we work on the lattice the energy-momentum relation is $\epsilon(k)=[(h-\cos(k))^{2}+\sin^{2}(k)]^{1/2}$. 
Once you take the proper scaling limit $a \to 0$, then $\epsilon(k)$ becomes relativistic and you can replace the general prefactor with a Bessel function $K_0$.
Comparisons between theory and numerical data are shown in figures (\ref{Fig:numerics1}) and (\ref{Fig:numerics2}). We note that in expression (\ref{eq:75}) $T$ is arbitrary (with the constraint $T>t>0$), and not necessarily much longer than the Fermi time $t_f=x/v_{max}$, as was in (\ref{eq:74tris}).
For larger quenches the semiclassical prediction, with  $f_k$ equal to $|K(k)|^{2}/(1+|K(k)|^{2})$ is not accurate. Indeed one has to take into account that for arbitrary values of the magnetic field $h>1$ generic excitations are no longer single spin-flips (even if this is not the only source of error, see below).
In general their shape is given by a superposition of states with an arbitrary number of spin-flips, with coefficients depending on the value of $h$ itself. 
Only when $h\gg 1$ are single excitations well-approximated by single-spin flips, as you treat them in the semi-classical approach. 
A first correction to the semiclassical results corresponds to substituting $f_k \to -1/2 \log|\cos(\Delta_k)|$, where $\Delta_k$ is the Bogoliubov angle given by: 
\begin{equation}
\label{eq:77bis}
\cos(\Delta_k)={\displaystyle\frac{h_0 h-(h_0 +h)\cos(k)+1}{\epsilon_{h_0}(k)\epsilon_{h}(k)}}.
\end{equation}
This substitution, which follows from asymptotically exact techniques (see \cite{Calabrese2}), increases the agreement between the theoretical predictions and numerical data, expecially close to the quantum critical point $h=1$ where excitations are no longer localized objects. 
However in some plots mismatch between theory and numerics is still present, and we believe that it is due to the hypothesis of 
particle-number conservation of the {\em pure} semi-classical approach (i.e., the number of quasiparticles is conserved during the dynamical evolution of any configuration). If we considered the possibility of creating or destroying particles during the time evolution (which is realized by an operator insertion), we would get time-dependent corrections to the prefactor (\ref{eq:76bis}). Roughly speaking this would mean going beyond the leading order in the semi-classical approximation, a topic that is not addressed in detail in this paper. 
In the recent article \cite{Evangelisti} the dynamic correlators after a quantum quench were studied by using the {\em form-factor} technique, finding that the prefactor of expression (\ref{eq:75}) can be written as:   
\be
\label{eq:exact1}
C^{\rm Ising}_{T=0}(x, t) \propto {\displaystyle\int_{-\pi}^{\pi}\frac{dk}{2\pi}\,\frac{{\rm e}^{ikx}}{\epsilon(k)}\left[{\rm e}^{-i\epsilon_{k}t}+2i\tan(\frac{\Delta_k}{2})\cos(\epsilon_k(T)){\rm sgn}(x-\epsilon'_k t)\right]}.
\ee
{\em Outside} the light-cone (when $x>v_{max}t$) the first contribution in (\ref{eq:exact1}) is exponentially small, whereas the second one behaves as a power law. This observation explains in turn why the {\em pure} semi-classical approximation typically fails when applied to this regime. In particular the semi-classics is not able to capture the behaviour of the equal-time correlator, that represents the extreme out-of-the light-cone case. \\

\subsection{The ferromagnetic phase}  %
\label{sec:subIsing2}

\begin{figure}[t]
\begin{center}
\includegraphics[width=7.5cm]{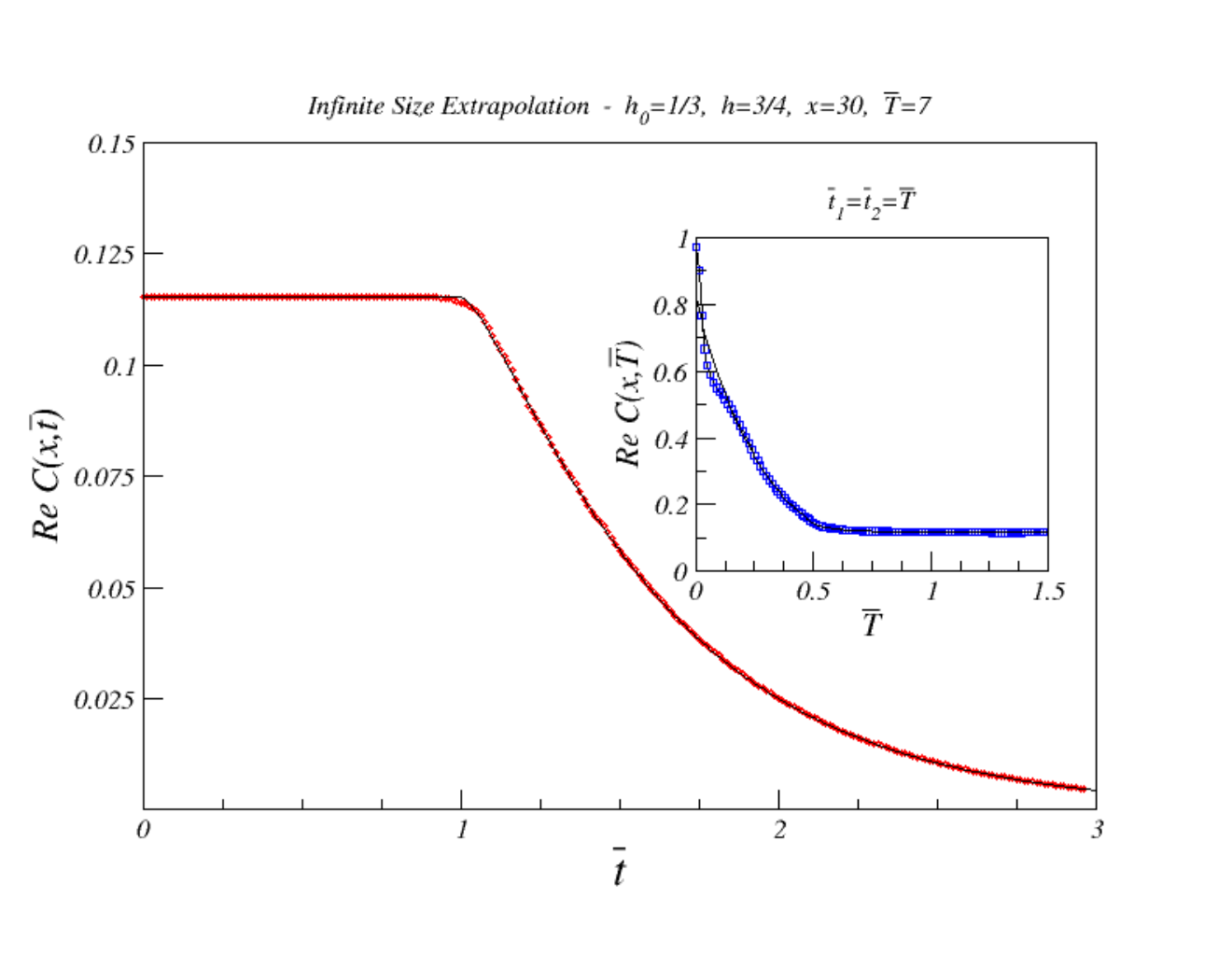}
\includegraphics[width=7.5cm]{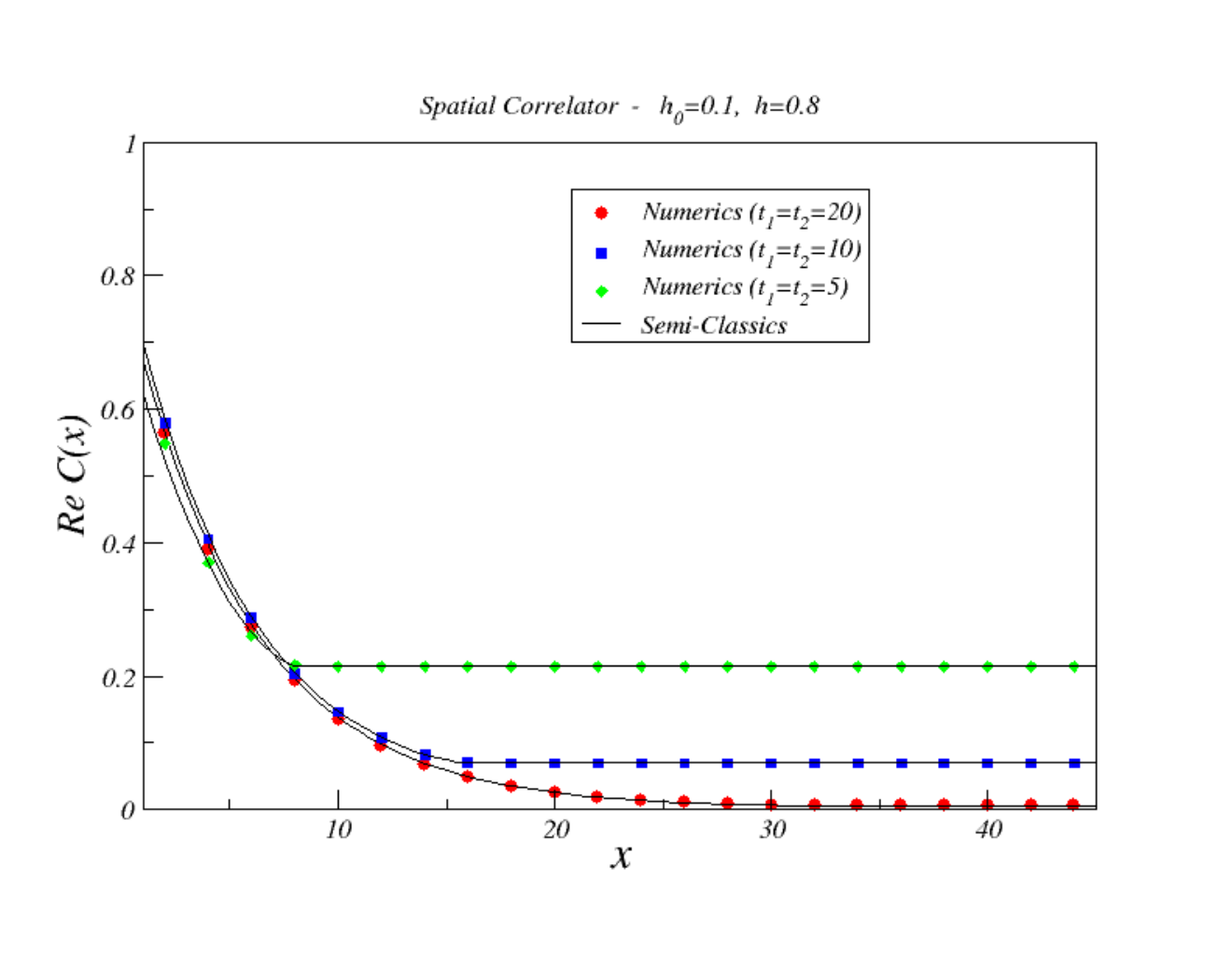}
\end{center}
\caption{Left panel (inset): real part of the non-equal-time (equal-time) two point function in the ordered phase. Again the numerical data  are extrapolated in the thermodynamic limit ($L \to \infty$). In this phase, unlike in the paramagnetic one, the agreement between the semi-classics and the numerics works well also for the equal-time correlator. Right panel: spatial correlation at equal times.}
\label{Fig:numerics3}
\end{figure}

Formula (\ref{eq:74tris}) cannot be directly used to derive the correlator of the Ising model in the ferromagnetic phase. In this case excitations are no longer spin-flips, in fact when $h<1$ these are domain walls (kinks). Following the general approach of F. Igloi and H. Reiger \cite{Igloi1,Igloi2} we can easily obtain the two-point correlation function for arbitrary value of times and distances. The key ideas of this method have already been extensively discussed in their works (even if these authors did not consider explicitly the case of dynamical correlators), here we just show the final result:
\be
\label{eq:ferr_phase}
C^{\rm Ising}(x, T, t)= C(h_0,h)\,R(x, T, t)
\ee
where the relaxation function in the ferromegnetic phase is equal to that in the paramagnetic phases
while the multiplicative constant, as introduced in \cite{Calabrese3,Calabrese4} reads as:
\be
\label{eq:constant1}
C(h_0,h)={\displaystyle\frac{1-hh_0+\sqrt{(1-h^2)(1-h_0^2)}}{2\sqrt{1-hh_0}\sqrt[4]{1-h_0^2}}}.
\ee
In addition it is worth noticing that the semi-classical two-point correlator in the ordered phase of the Ising model is always a real function, as was the dominant contribution of the form-factor result at large ($x,T$) in \cite{Evangelisti}. 
In the ferromagnetic phase  the equal-time correlators can also be described by the semi-classic approach, in contrast to the paramagnetic phase, as is shown in the inset of figure (\ref{Fig:numerics3}) 

\section{Conclusions}
\label{s:conclusions}
We have developed a semi-classical theory for the out-of-equilibrium quantum relaxation of the O(3) non-linear sigma model, after having prepared the system in a coherent superposition of Cooper pairs, a structure that is in agreement with the integrability of the theory in the bulk. For such quenches we analyzed the two-point function of the order parameter $\tilde{n}_{z}$ and argued that, in the long time limit its expression is given by formula (\ref{eq:74tris}), while for arbitrary times by equation (\ref{eq:74quatris}). The method employed here is a generalization of that used for studying the finite-temperature behaviour of a series of one-dimensional chains \cite{Rapp2,Sachdev,Rapp1}. As was already observed in other integrable models, the long-time behaviour of this two-point function after a quantum quench is {\em not} thermal. 
For equal times ($t_2=t_1\gg x/v_{max}$) the relaxation (\ref{eq:74tris}) reaches a stationary state, while for non-equal times ($t_2\neq t_1$) it is expressed as a function of the time difference $t$. These features are strong indications that the long-time limit of the two-point function (\ref{eq:5}) can be described by a statistical ensemble, and we expect it to be the GGE, being the model integrable.
It would be interesting to have an independent and explicit result in the GGE framework, in order to check the general belief also for the O(3) non-linear sigma model. The main difficulty here consists in finding an analytical expression for the lagrangian parameters of the GGE Hamiltonian, in order to explicitly carry out the statistical average.\\
The semi-classical approach has also been applied to predict the dynamics of the order-parameter two-point function of the transverse field Ising chain, in both phases.
In this case results for the auto-correlation and for the equal-time correlation are already present in literature \cite{Igloi4,Igloi1}, the novelty here is the generalization of the semi-classical method to the case of different-times correlators. Yet, the exact results recently obtained in \cite{Evangelisti}, have allowed us to explore the limit of the {\em pure} semi-classics. This technique works very well within the ferromagnetic phase (also for pretty large quenches), whilst it is in the paramagnetic phase where it shows its real limits: it fails in predicting the equal-time two-point correlation functions, and in general the agreement in the out-of-the-light-cone case (see the insets of figures (\ref{Fig:numerics1}) and  (\ref{Fig:numerics2})) is only qualitative. Furthermore in the proximity of the critical point, the method - which in general is supposed to work well only for small quenches - needs to be improved with the substitution $f_k \to -1/2 \log|\cos(\Delta_k)|$, which takes into account the real shape of the excitations.\\
Nevertheless, our semi-classical theory can be applied straightforwardly to several other models, integrable and non-integrable, including the q-Potts model and the sine-Gordon theory, in order to compute correlators. These models will generate different combinatorial problems (the nature of the excitations is, of course, model-dependent), but the main ideas remain the same. 
\ack
I am extremely grateful to Fabian Essler for the suggestion of working on this problem and for all the instructive discussions we have had over the last year. Without his support this work would have never seen the light of day. I am also in great debt to Maurizio Fagotti for his priceless help and for the huge number of useful hints he gave me.\\
Finally I want to thank N.J. Robinson, J. Cardy, A.M. Tsvelik, R.M. Konik and S. Sotiriadis  for fruitful and pleasant discussions.

\appendix
\setcounter{section}{0}

\section{Occupation numbers and mode probabilities}  %
\label{app:occ_numb}
In this appendix we compute the occupation numbers and the probability distributions for each mode, starting from the definition of the squeezed coherent state (\ref{eq:2}) and the operator algebra of the operators $(Z, Z^{\dag})$. Let us recall the definition of the initial state $|\psi\rangle$:
\begin{equation}
\label{eq:76}
|\psi\rangle = \exp{\left (\sum_{a, b}\int_{0}^{\infty} \frac{dk}{2\pi}\, K^{ab}(k)Z^{\dag}_{a}(-k)Z^{\dag}_{b}(k) \right)}|0\rangle,
\end{equation}
The $Z$ operators obey the Zamolodchikov-Faddeev (ZF) algebra, namely:
\begin{equation}
\label{eq:77}
\begin{array}{l}
Z_{a}(k_1)Z_{b}(k_2)-\mathcal{S}_{ab}^{cd}(k_1, k_2)Z_{d}(k_2)Z_{c}(k_1)=0\\
Z_{a}^{\dag}(k_1)Z_{b}^{\dag}(k_2)-\mathcal{S}_{ab}^{cd}(k_1, k_2)Z_{d}^{\dag}(k_2)Z_{c}^{\dag}(k_1)=0\\
Z_{a}(k_1)Z_{b}^{\dag}(k_2)-\mathcal{S}_{cb}^{ad}(k_1, k_2)Z_{d}^{\dag}(k_2)Z_{c}(k_1)=\delta_{ab}\,\delta(k_1-k_2).
\end{array}
\end{equation}
Intuitively this means that the exchange of two quasiparticle is realized by the two-particle scattering matrix $\mathcal{S}(k_1, k_2)$. In the present case we assume the form (\ref{eq:6}) for the $\mathcal{S}$ matrix, and thus the ZF algebra becomes: 
\begin{equation}
\label{eq:78}
\begin{array}{l}
Z_{a}(k_1)Z_{b}(k_2)+Z_{a}(k_2)Z_{b}(k_1)=0\\
Z_{a}^{\dag}(k_1)Z_{b}^{\dag}(k_2)+Z_{a}^{\dag}(k_2)Z_{b}^{\dag}(k_1)=0\\
Z_{a}(k_1)Z_{b}^{\dag}(k_2)=[-\sum_{c}Z_{c}^{\dag}(k_2)Z_{c}(k_1)+\delta(k_1-k_2)]\delta_{ab}.
\end{array}
\end{equation}
The ground state $|0\rangle$ satisfies $Z_{a}(k)|0\rangle =0$, $\forall{a}$ $\forall{k}$. From equations (\ref{eq:77}) it is clear that this algebra encodes the fact that the order of quantum number is preserved in time. The main quantities we want to compute are the occupation numbers $f_{k}^{ab}$, which are defined by:
\begin{equation}
\label{eq:79}
\frac{\langle\psi|Z_{a}^{\dag}(-k)Z_{b}^{\dag}(k)Z_{b}(k)Z_{a}(-k)|\psi\rangle}{\langle\psi|\psi\rangle}\equiv f_{k}^{ab}.
\end{equation}
This quantity is exactly the number of quasiparticle pairs of kind $(a, b)$ with momenta $(-k, k)$ (see figure (\ref{fig:5})), and it is directly connected to the probabilities $P^{\lambda\lambda'}$ we defined in the body of the paper. 
The idea is to treat the $K$-matrix as an expansion parameter, but working from the beginning in the thermodynamic limit the squeezed state does not have a good expansion. Divergences appear in the expansion terms, in the form of square Dirac delta-functions. In general there are two different methods of regularizing these divergences: one directly regulates the integral expressions in the infinite volume whereas the other operates through subtracting divergences in a large, finite volume. These two techniques were proposed and compared in \cite{Fabian1}. In the follow we shall use the second method.\\
Let us start by analysing the expansion of the denominator of (\ref{eq:79}). First Taylor expand the squezeed state as:
\begin{equation}
\label{eq:84}
|\psi\rangle= (1+ \sum_{\stackrel {k>0}{a, b}}K^{ab}(k)Z^{\dag}_{a}(-k)Z^{\dag}_{b}(k)  + \frac{1}{2!}(\sum_{\stackrel {k>0}{a, b}}K^{ab}(k)Z^{\dag}_{a}(-k)Z^{\dag}_{b}(k))^{2}+\dots)|0\rangle,
\end{equation}
therefore we have:
\begin{equation}
\label{eq:85}
\langle\psi|\psi\rangle=1+Z_1+Z_2+\dots,
\end{equation}
where 
\begin{equation}
\label{eq:86}
Z_1= \sum_{\stackrel {k>0}{a, b}}\sum_{\stackrel {\xi>0}{c, d}} (K^{ab}(k))^{*}K^{cd}(\xi)\langle 0| Z_{b}(k)Z_{a}(-k)Z_{c}^{\dag}(-\xi)Z_{d}^{\dag}(\xi)|0\rangle = \sum_{\stackrel {k>0}{a, b}}|K^{ab}(k)|^{2}.
\end{equation} 
As expected this term is proportional to the volume $L$. the higher orders will be proportional to $L^{2}$, $L^{3}$ {\em et cetera}. Let us consider the expansion of the numerator of (\ref{eq:79}):
\begin{equation}
\label{eq:87}
\langle\psi| Z_{\alpha}^{\dag}(-\xi)Z_{\beta}^{\dag}(\xi)Z_{\beta}(\xi)Z_{\alpha}(-\xi)|\psi\rangle = W_1+\frac{1}{4} W_2+\dots,
\end{equation}
where the first term $W_1$is given by:
\begin{equation}
\label{eq:88}
\fl W_1=\sum_{\stackrel {k_1>0}{a, b}}\sum_{\stackrel {k_2>0}{c, d}}\langle 0| Z_{c}(k_2)Z_{d}(-k_2)Z_{\alpha}^{\dag}(-\xi)Z_{\beta}^{\dag}(\xi)Z_{\beta}(\xi)Z_{\alpha}(-\xi)Z_{a}^{\dag}(-k_1)Z_{b}^{\dag}(k_1)|0\rangle = |K^{\alpha\beta}(\xi)|^{2}.
\end{equation}
This quantity is finite, also in the infinite volume limit $L\to\infty$, whilst $W_2$ is proportional to $L$. We now need the expression of $W_2$ to see how these divergences cancel against the normalization of the boundary state. By definition we have:
\begin{equation}
\label{eq:89}
\begin{array}{l}
 W_2 = {\displaystyle\sum_{\stackrel {k_1>0}{a, b}}\sum_{\stackrel {k_2>0}{c, d}}\sum_{\stackrel {k_3>0}{e, f}}\sum_{\stackrel {k_4>0}{g, h}}\langle 0| Z_{a}(k_1)Z_{b}(-k_1)Z_{c}(k_2)Z_{d}(-k_2)Z_{\alpha}^{\dag}(-\xi)Z_{\beta}^{\dag}(\xi)}\\
 {\displaystyle \times Z_{\beta}(\xi)Z_{\alpha}(-\xi)Z_{e}^{\dag}(-k_3)Z_{f}^{\dag}(k_3)Z_{g}^{\dag}(-k_4)Z_{h}^{\dag}(k_4)|0\rangle}
\end{array}
\end{equation}
After a bit of algebra we end up with the following result for $W_2$:
\begin{equation}
\label{eq:89bis}
 W_2 =  {\displaystyle 4|K^{\alpha\beta}(\xi)|^{2}\sum_{\stackrel {k>0}{a, b}}|K^{ab}(k)|^{2}+2|K^{\alpha\beta}(\xi)|^{2}\sum_{ a, b}|K^{ab}(\xi)|^{2}},
\end{equation}
where we immediately see that the first term is of order $L$. In particular the first term of (\ref{eq:89}) can be written as $4 W_1 Z_1$, thus the first correction in equation (\ref{eq:79}) is given by:
\begin{equation}
\label{eq:90}
\frac{\langle\psi|Z_{a}^{\dag}(-k)Z_{b}^{\dag}(k)Z_{b}(k)Z_{a}(-k)|\psi\rangle}{\langle\psi|\psi\rangle}={\displaystyle\frac{W_1(1+Z_1)}{1+Z_1}}+\dots=W_1 + o(W_1).
\end{equation}
We see that the divergent term indeed cancels against the normalization factor, and this happens order by order. We conclude that as long as $K^{ab}(k)$ is small we can calculate the matrix elements like (\ref{eq:79}) by expanding the squeezed state. The final result can be written as:
\begin{equation}
\label{eq:80}
\boxed{
f_{k}^{ab}=|K^{ab}(k)|^{2}+o(|K^{ab}(k)|^{2})}
\end{equation}
which explains the name {\em amplitudes} for the matrix elements $K^{ab}(k)$. The momentum occupation numbers $f_k$ are given by:
\begin{equation}
\label{eq:81}
f_{k}= \sum_{a,b}f_{k}^{ab},
\end{equation}
while for the probability densities $P^{ab}(k)$ -which are defined in the thermodynamic limit- we have:
\begin{equation}
\label{eq:82}
P^{ab}(k)=\frac{{\displaystyle f_{k}^{ab}}}{{\displaystyle\sum_{a,b}\int_{0}^{\infty}\frac{dk}{2\pi}f_{k}^{ab}}},
\end{equation}
which makes the normalization condition $\sum_{a,b}\int_{0}^{\infty}\frac{dk}{2\pi}P^{ab}(k)=1$ explicit. The probability $P^{ab}$ is defined by $P^{ab}=\int_{0}^{\infty}\frac{dk}{2\pi}P^{ab}(k)$.

\section{General expression of the relaxation function $R(r, t, T)$}
\label{app:general}
In this appendix we show the form of the relaxation function for the O(3) non-linear sigma model when the times $T$ and $t$ are arbitrary, with the only constraint $T>t>0$. 
This expression can be written as: 
\be
\label{eq:74quatris}
 R(r; t, T)  =  R_{1}(r; t, T) + R_{2}(r; t, T),
\ee
where the first addend reads as:
\begin{equation}
\label{eq:74penta}
\begin{array}{l}
\fl\quad R_{1}(r; t, T)={\displaystyle\int_{-\pi}^{\pi}\frac{d\phi}{2\pi}}\exp{\left(-t(1-\cos\phi){\displaystyle\int_{0}^{\infty}\frac{dk}{\pi}f_k\, v_{k}\Theta[v_{k}t-r]}\right)} \\
\fl\quad\times \exp{\left(-T(1+\cos2\phi-2\cos\phi){\displaystyle\int_{0}^{\infty}\frac{dk}{2\pi}f_k v_k\,\Theta[r-v_{k}T]}\right)}\\
\fl\quad\times\exp{\left(-r(1-\cos\phi){\displaystyle\int_{0}^{\infty}\frac{dk}{\pi}f_k\,\Theta[r-v_{k}t]\Theta[v_k T-r]}\right)} \\
\fl\quad\times\exp{\left(-r(1-\cos2\phi){\displaystyle\int_{0}^{\infty}\frac{dk}{2\pi}f_k\,\Theta[r-v_{k}T]}\right)}\\
\fl\quad\times\cos\left\{r\sin2\phi{\displaystyle\int_{0}^{\infty}\frac{dk}{2\pi}f_k\Theta[r-v_k T] + r\sin\phi\int_{0}^{\infty}\frac{dk}{\pi}f_k\Theta[v_k T-r]}\right. \\
\fl\quad+\left.2T(\sin\phi-\sin2\phi) {\displaystyle\int_{0}^{\infty}\frac{dk}{2\pi}f_k v_k\,\Theta[r-v_k T]}\right\}\\
\fl\quad\times{\displaystyle\frac{1+2\cos(\phi)(P^{00}-1)Q+\cos(2\phi)(Q^{2}-P^{00})-P^{00}Q^{2}}{1-2P^{00}\cos(2\phi)+(P^{00})^{2}}}\\,
\end{array}
\end{equation}
and the second term is given by
\begin{equation}
\label{eq:74pentabis}
\begin{array}{l}
\fl\quad R_{2}(r; t, T)={\displaystyle\int_{-\pi}^{\pi}\frac{d\phi}{2\pi}}\exp{\left(-2(T-t\cos\phi){\displaystyle\int_{0}^{\infty}\frac{dk}{2\pi}f_k\, v_{k}\Theta[v_{k}T-r]}\right)}  \\
\fl\quad\times\exp{\left(-r(1-\cos2\phi){\displaystyle\int_{0}^{\infty}\frac{dk}{2\pi}f_k\,\Theta[r-v_{k}T]}\right)}\\
\fl\quad\times\exp{\left([-T(1+\cos2\phi) +2t\cos\phi]{\displaystyle\int_{0}^{\infty}\frac{dk}{2\pi}f_k v_k\,\Theta[r-v_{k}T]}\right)}\\
\fl\quad\times\cos\left\{2r\sin\phi{\displaystyle\int_{0}^{\infty}\frac{dk}{2\pi}f_k\Theta[v_k t -r]+2t\sin\phi\int_{0}^{\infty}\frac{dk}{2\pi}f_k v_k\Theta[r-v_k t]\Theta[v_k T -r]} \right.\\
\fl\quad+ \left. (-T\sin2\phi+t\sin\phi){\displaystyle\int_{0}^{\infty}\frac{dk}{2\pi}f_k v_k\,\Theta[r-v_k T]+r\sin2\phi\int_{0}^{\infty}\frac{dk}{2\pi}f_k\,\Theta[r-v_k T]}\right\}\\
\fl\quad\times {\displaystyle\frac{P^{00}Q^{2}-(P^{00})^{2}-\cos(2\phi)(Q^{2}-P^{00})}{1-2P^{00}\cos(2\phi)+(P^{00})^{2}}}.
\end{array}
\end{equation}
This formula represents the leading order of the semi-classical approach to the $n_z-n_z$ correlator in the $O(3)$ non-linear sigma model. From this expression all subcases can be derived straightforwardly. 
\section*{References}


\end{document}